\documentclass[aps,physrev,10pt,floatfix,longbibliography,nofootinbib,reprint,superscriptaddress]{revtex4-2}
\usepackage{amssymb}
\usepackage[english]{babel}
\usepackage{bbm}
\usepackage{bm}
\usepackage{graphicx}
\usepackage{hyperref}
\usepackage[notrig]{physics}
\usepackage{pifont}

\newcommand{\bbone}{\ensuremath{\mathbbm{1}}}
\newcommand{\JoneJtwo}{$J_1$-$J_2$}
\newcommand{\llangle}{{\langle \langle}}
\newcommand{\rrangle}{{\rangle \rangle}}
\newcommand{\spindown}{{\downarrow}}
\newcommand{\spinup}{{\uparrow}}

\DeclareMathOperator{\Var}{Var}

\newcommand{\Eq}[1]{Eq.~\eqref{eq:#1}}

\begin{document}

\title{Variational Benchmarks for Quantum Many-Body Problems}
\author{Dian Wu}
\affiliation{Institute of Physics, \'Ecole Polytechnique F\'ed\'erale de Lausanne (EPFL), CH-1015 Lausanne, Switzerland}
\affiliation{Center for Quantum Science and Engineering, \'Ecole Polytechnique F\'ed\'erale de Lausanne (EPFL), CH-1015 Lausanne, Switzerland}
\author{Riccardo Rossi}
\affiliation{Institute of Physics, \'Ecole Polytechnique F\'ed\'erale de Lausanne (EPFL), CH-1015 Lausanne, Switzerland}
\affiliation{Sorbonne Universit\'e, CNRS, Laboratoire de Physique Th\'eorique de la Mati\`ere Condens\'ee (LPTMC), F-75005 Paris, France}
\author{Filippo Vicentini}
\affiliation{Center for Quantum Science and Engineering, \'Ecole Polytechnique F\'ed\'erale de Lausanne (EPFL), CH-1015 Lausanne, Switzerland}
\affiliation{CPHT, CNRS, \'Ecole Polytechnique, Institut Polytechnique de Paris, F-91128 Palaiseau, France}
\affiliation{Coll\`ege de France, Universit\'e PSL, F-75005 Paris, France}
\author{Nikita Astrakhantsev}
\affiliation{Department of Physics, University of Zurich, CH-8057 Zurich, Switzerland}
\author{Federico Becca}
\affiliation{Dipartimento di Fisica, Universit\`a di Trieste, I-34151 Trieste, Italy}
\author{Xiaodong Cao}
\affiliation{Center for Computational Quantum Physics, Flatiron Institute, New York, New York 10010, USA}
\author{Juan Carrasquilla}
\affiliation{Vector Institute, MaRS Centre, Toronto, Ontario M5G 1M1, Canada}
\affiliation{Institute for Theoretical Physics, ETH Z\"urich, CH-8093 Zurich, Switzerland}
\author{Francesco Ferrari}
\affiliation{Institut f\"ur Theoretische Physik, Goethe-Universit\"at, 60438 Frankfurt am Main, Germany}
\author{Antoine Georges}
\affiliation{Coll\`ege de France, Universit\'e PSL, F-75005 Paris, France}
\affiliation{Center for Computational Quantum Physics, Flatiron Institute, New York, New York 10010, USA}
\affiliation{CPHT, CNRS, \'Ecole Polytechnique, Institut Polytechnique de Paris, F-91128 Palaiseau, France}
\affiliation{Department of Quantum Matter Physics, Universit\'e de Gen\`eve, CH-1211 Geneva, Switzerland}
\author{Mohamed Hibat-Allah}
\affiliation{Vector Institute, MaRS Centre, Toronto, Ontario M5G 1M1, Canada}
\affiliation{Perimeter Institute for Theoretical Physics, Waterloo, Ontario N2L 2Y5, Canada}
\affiliation{Department of Physics and Astronomy, University of Waterloo, Waterloo, Ontario N2L 3G1, Canada}
\affiliation{Department of Applied Mathematics, University of Waterloo, Waterloo, Ontario N2L 3G1, Canada}
\author{Masatoshi Imada}
\affiliation{Toyota Physical and Chemical Research Institute, Nagakute, Aichi 480-1192, Japan}
\affiliation{Waseda Research Institute for Science and Engineering, Waseda University, Shinjuku-ku, Tokyo 169-8555, Japan}
\affiliation{Physics Division, Sophia University, Chiyoda-ku, Tokyo 102-8554, Japan}
\affiliation{Department of Applied Physics, University of Tokyo, Bunkyo-ku, Tokyo 113-8656, Japan}
\author{Andreas M. Läuchli}
\affiliation{Laboratory for Theoretical and Computational Physics, Paul Scherrer Institute, CH-5232 Villigen, Switzerland}
\affiliation{Institute of Physics, \'Ecole Polytechnique F\'ed\'erale de Lausanne (EPFL), CH-1015 Lausanne, Switzerland}
\author{Guglielmo Mazzola}
\affiliation{Institute for Computational Science, University of Zurich, CH-8057 Zurich, Switzerland}
\author{Antonio Mezzacapo}
\affiliation{IBM Quantum, T. J. Watson Research Center, Yorktown Heights, New York 10598, USA}
\author{Andrew Millis}
\affiliation{Department of Physics, Columbia University, New York, New York 10027, USA}
\affiliation{Center for Computational Quantum Physics, Flatiron Institute, New York, New York 10010, USA}
\author{Javier Robledo Moreno}
\affiliation{Center for Computational Quantum Physics, Flatiron Institute, New York, New York 10010, USA}
\affiliation{Center for Quantum Phenomena, Department of Physics, New York University, New York, New York 10003, USA}
\author{Titus Neupert}
\affiliation{Department of Physics, University of Zurich, CH-8057 Zurich, Switzerland}
\author{Yusuke Nomura}
\affiliation{Department of Applied Physics and Physico-Informatics, Keio University, Kohoku-ku, Yokohama 223-8522, Japan}
\affiliation{Institute for Materials Research, Tohoku University, Aoba-ku, Sendai 980-8577, Japan}
\author{Jannes Nys}
\affiliation{Institute of Physics, \'Ecole Polytechnique F\'ed\'erale de Lausanne (EPFL), CH-1015 Lausanne, Switzerland}
\affiliation{Center for Quantum Science and Engineering, \'Ecole Polytechnique F\'ed\'erale de Lausanne (EPFL), CH-1015 Lausanne, Switzerland}
\author{Olivier Parcollet}
\affiliation{Center for Computational Quantum Physics, Flatiron Institute, New York, New York 10010, USA}
\affiliation{Universit\'e Paris-Saclay, CNRS, CEA, Institut de physique th\'eorique, F-91191 Gif-sur-Yvette, France}
\author{Rico Pohle}
\affiliation{Waseda Research Institute for Science and Engineering, Waseda University, Shinjuku-ku, Tokyo 169-8555, Japan}
\affiliation{Department of Applied Physics, University of Tokyo, Bunkyo-ku, Tokyo 113-8656, Japan}
\author{Imelda Romero}
\affiliation{Institute of Physics, \'Ecole Polytechnique F\'ed\'erale de Lausanne (EPFL), CH-1015 Lausanne, Switzerland}
\affiliation{Center for Quantum Science and Engineering, \'Ecole Polytechnique F\'ed\'erale de Lausanne (EPFL), CH-1015 Lausanne, Switzerland}
\author{Michael Schmid}
\affiliation{Waseda Research Institute for Science and Engineering, Waseda University, Shinjuku-ku, Tokyo 169-8555, Japan}
\author{J. Maxwell Silvester}
\affiliation{Department of Physics and Astronomy, University of California, Irvine, California 92697, USA}
\author{Sandro Sorella}
\affiliation{SISSA, International School for Advanced Studies, I-34136 Trieste, Italy}
\author{Luca F. Tocchio}
\affiliation{Institute for Condensed Matter Physics and Complex Systems, Department of Applied Science and Technology (DISAT), Politecnico di Torino, I-10129 Torino, Italy}
\author{Lei Wang}
\affiliation{Institute of Physics, Chinese Academy of Sciences, Beijing 100190, China}
\affiliation{Songshan Lake Materials Laboratory, Dongguan, Guangdong 523808, China}
\author{Steven R. White}
\affiliation{Department of Physics and Astronomy, University of California, Irvine, California 92697, USA}
\author{Alexander Wietek}
\affiliation{Max Planck Institute for the Physics of Complex Systems, 01187 Dresden, Germany}
\author{Qi Yang}
\affiliation{Institute of Physics, Chinese Academy of Sciences, Beijing 100190, China}
\affiliation{University of Chinese Academy of Sciences, Beijing 100049, China}
\author{Yiqi Yang}
\affiliation{Department of Physics, College of William and Mary, Williamsburg, Virginia 23187, USA}
\author{Shiwei Zhang}
\affiliation{Center for Computational Quantum Physics, Flatiron Institute, New York, New York 10010, USA}
\author{Giuseppe Carleo}
\email{Corresponding author. Email: giuseppe.carleo@epfl.ch}
\affiliation{Institute of Physics, \'Ecole Polytechnique F\'ed\'erale de Lausanne (EPFL), CH-1015 Lausanne, Switzerland}
\affiliation{Center for Quantum Science and Engineering, \'Ecole Polytechnique F\'ed\'erale de Lausanne (EPFL), CH-1015 Lausanne, Switzerland}
\date{Sept.\ 12, 2024}

\begin{abstract}
The continued development of computational approaches to many-body ground-state problems in physics and chemistry calls for a consistent way to assess its overall progress.
In this work, we introduce a metric of variational accuracy, the V-score, obtained from the variational energy and its variance.
We provide an extensive curated dataset of variational calculations of many-body quantum systems, identifying cases where state-of-the-art numerical approaches show limited accuracy, and future algorithms or computational platforms, such as quantum computing, could provide improved accuracy.
The V-score can be used as a metric to assess the progress of quantum variational methods toward a quantum advantage for ground-state problems, especially in regimes where classical verifiability is impossible.
\clearpage
\end{abstract}

\maketitle

\begin{figure*}[tb]
\includegraphics[width=\linewidth]{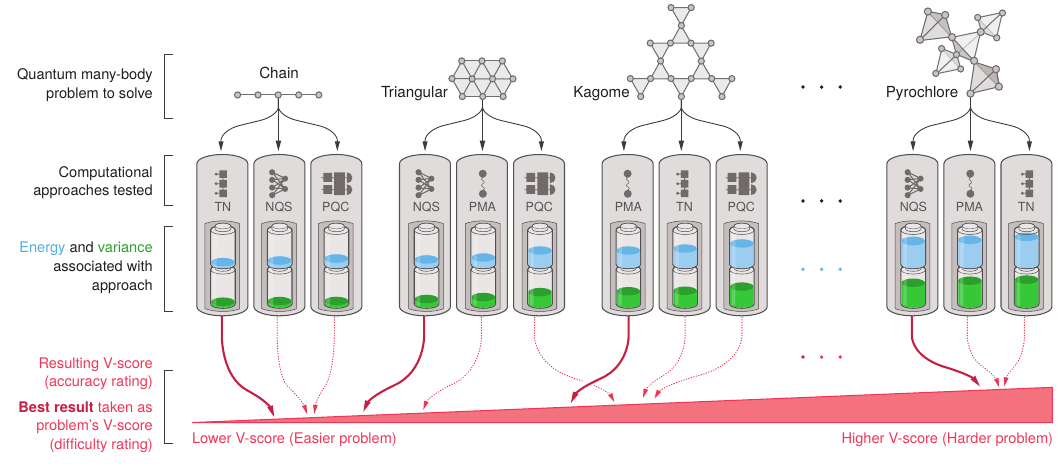}
\caption{\textbf{Sketch of the V-score as a metric of simulation hardness.}
In this work, we present an extensive dataset of computational results for quantum many-body ground-state problems (for this sketch, we have selected a spin-$\frac{1}{2}$ system on a chain, a kagome, and a pyrochlore lattice).
For each Hamiltonian in the dataset, we compute the mean energy and its variance with different variational techniques, including physically motivated ansatzes (PMAs), neural quantum states (NQSs), tensor networks (TNs), and parameterized quantum circuits (PQCs).
The energy and the variance are combined into the V-score, a metric of variational accuracy that we introduce in the main text. A low V-score is associated with high accuracy.
The best result per Hamiltonian is then taken as the V-score for that Hamiltonian, and used to rank the Hamiltonians in terms of simulation accuracy, highlighting which quantum many-body models are hard to simulate with current methods.
}
\label{fig:sketch}
\end{figure*}

\section{Introduction}

A key aspect of the quantum many-body problem, for systems ranging from the subatomic to molecules and materials, is determining the ground-state properties and energy. Knowing the ground state, one can predict which systems are stable and whether these systems exhibit useful and exotic phases, such as superconductivity or spin liquid states.
However, because of the exponential complexity of the quantum wave function, finding the ground state of a many-body system can be very challenging, which limits exact numerical studies to a small number of particles. Efficiently solving the general ground-state problem is largely believed to be intractable. However, this does not necessarily apply to any particular system or class of systems, which may admit powerful approximations for ground states. Decades of research have focused on devising computational methods to find approximate solutions for specific cases of interest.

These computational methods have widely varying degrees of accuracy, and typically each method is much more successful on some systems than on others. Some of the most widely used methods include quantum Monte Carlo (QMC)~\cite{ceperley_quantum_1986,becca_sorella_2017,Shiwei_LectureNotes2019}, tensor networks (TNs)~\cite{white_density_1992,orusTensorNetworksComplex2019}, and dynamical mean field theory (DMFT) and its extensions~\cite{georges1992hubbard,georges_dmft_1996}. It is known that the applicability of QMC methods is negatively affected by the frustration of the quantum system and particle statistics~\cite{troyer_computational_2005}; similarly, high entanglement and large correlation lengths limit the applicability of TN~\cite{cirac_matrix_2021} and DMFT~\cite{georges_dmft_1996}, respectively. Variational approaches based on physically motivated ansatzes~\cite{McMillan1965,Ceperley1977} or neural networks~\cite{carleo_solving_2017} are not explicitly affected by the aforementioned issues. However, it is more difficult to assess their applicability and accuracy for a given quantum many-body system.

Quantum computers provide an alternative platform to attack quantum many-body problems~\cite{feynman_simulating_1982}. Notably, the dynamics of quantum many-body systems can be efficiently simulated by a digital quantum computer when the initial states are easy to prepare~\cite{lloyd_universal_1996}. Besides dynamics, substantial attention has been devoted to preparing ground states that are difficult to study with classical algorithms. Quantum algorithms for this task include phase estimation~\cite{kitaev_quantum_1995}, variational approaches~\cite{Peruzzo_2014,kandala_hardware-efficient_2017,cerezo2021variational,bharti_noisy_2022}, adiabatic passage~\cite{farhi2000adiabatic}, imaginary time evolution~\cite{motta2020qite_qlanczos}, and subspace and Lanczos methods~\cite{parrish2019filterdiagonalization,kirby2022exact}.

A fundamental challenge in assessing newly established computational methods based on classical or quantum computing is defining a consistent accuracy metric. Especially for ground-state problems, such a metric is necessary to clearly identify target Hamiltonians of broad interest, which cannot be solved with sufficient accuracy by existing methods.
Also, this metric is crucial to quantify the improvements of computational approaches with time. In the context of assessing quantum computing-based methods, this issue pertains to the broader problem of determining in what cases quantum computers have an advantage over classical ones~\cite{bravyi_quantum_2018,bouland_complexity_2019}.

Determining a consistent metric for physically and chemically relevant ground-state problems is one of the goals of this work. To this end, we provide a large, curated collection of variational and numerically exact results on strongly correlated lattice models obtained by both state-of-the-art and baseline methods.
The data that we provide include multiple approaches, such as exact diagonalization (ED), QMC~\cite{ceperley_quantum_1986} in the auxiliary field algorithm~\cite{Blankenbecler_1981,Sorella_1989,Imada_1989,Shiwei_PRB1997,Hao_PRA2015}, matrix product states (MPSs)~\cite{white_density_1992}, variational wave functions formulated on a lattice~\cite{Horsch1983}, and neural network-based methods~\cite{carleo_solving_2017}.
In addition to providing the data, we introduce an indicator of the variational accuracy of these results, named the V-score, that is suitable for directly comparing classical and quantum computing-based variational approaches.
The V-score, obtained as a combination of the mean energy and its variance of a given variational state, allows us to identify what Hamiltonians and regimes are hard to approximate with classical variational methods without prior knowledge of the exact solution.
Furthermore, we argue that the V-score can be used as a controlled benchmark to quantify the continued progress of quantum algorithms and quantum hardware to simulate those challenging target Hamiltonians.

\section{Benchmarking variational algorithms}

We focus our study on benchmarking classical and quantum variational algorithms in approximating ground states of quantum many-body systems. On the classical side, these algorithms involve explicitly maintained variational representations of wave functions, such as TN or variational Monte Carlo (VMC)-based approaches.
On the quantum side, the variational methods of major interest involve parameterized quantum circuits (PQCs) or other state preparation techniques based on local unitary transformations. In all cases, we assume that the methods to be benchmarked allow unbiased estimates of expectation values for Hamiltonians with few-body interactions ($k$-local operators, in the language of quantum information). Such expectation values can be obtained with a controllable statistical error, as in the case of classical Monte Carlo-based techniques, or as a result of statistical noise caused by measurements on quantum hardware.

\subsection{Choice of problems}

There is large freedom in the choice of many-body quantum problems that can be used to benchmark computational techniques. In this work, we have decided to focus on lattice Hamiltonians. These are minimal models of strong correlations and typically capture the essence of many physical systems.
Lattice models first rose to prominence within classical statistical mechanics with the definition of the Ising model~\cite{nissHistoryLenzIsing2008}. Within solid-state physics, they stem from tight binding approaches to describe the electronic band structure~\cite{PhysRev.94.1498}. More recently, within the second quantization formalism, they are routinely used in different areas of physics to understand the low-energy behavior of unconventional quantum phases and transitions among them~\cite{sachdev2001quantum,wen2004quantum}. In this regard, the transverse-field Ising model (TFIM) provides the simplest example of a zero-temperature phase transition purely driven by quantum fluctuations between a paramagnet and a ferromagnet as seen, for example, in the Ising ferromagnet LiHoF$_4$~\cite{PhysRevLett.77.940,sachdev2001quantum}.
Other prominent examples are the various quantum impurity models, in which a localized interacting degree of freedom is embedded into a non-interacting bulk, such as the Anderson impurity model~\cite{PhysRev.124.41}.
Quantum impurity models are central to quantum embedding methods, such as DMFT~\cite{georges_dmft_1996}, and have applications to nanoelectronic devices~\cite{Kouwenhoven_2001}. Their lattice generalizations, such as the Kondo lattice model, describe heavy fermion systems with 4f or 5f atoms, such as Ce or U~\cite{hewson_1993}.

Similarly, the Hubbard model~\cite{hubbardElectronCorrelationsNarrow1963,kanamoriElectronCorrelationFerromagnetism1963,PhysRevLett.10.159} has been widely used to capture the essence of strong correlation in solids and has been shown to be relevant to the study of high-temperature superconductivity in cuprate compounds---e.g., La$_{2-x}$Sr$_x$CuO$_4$~\cite{leeDopingMottInsulator2006}---and the Mott metal-insulator transitions in a variety of compounds~\cite{imadaMetalinsulatorTransitions1998}. A descendant of the Hubbard model, the Heisenberg model describes a wide range of magnetic phases, for example, with ferromagnetic or antiferromagnetic orders~\cite{RevModPhys.63.1}. In addition, when defined on geometrically frustrated lattices, possibly with anisotropic super-exchange couplings, the Heisenberg model gives rise to a wealth of phenomena, including spin liquid phases with topological order and exotic critical points~\cite{balentsSpinLiquidsFrustrated2010,savaryQuantumSpinLiquids2016}. In this respect, the rare earth compound YbMgGaO$_4$~\cite{PhysRevLett.115.167203} and the mineral herbertsmithite ZnCu$_3$(OH)$_6$Cl$_2$~\cite{RevModPhys.88.041002} have offered examples for unconventional quantum phases on triangular and kagome lattices.

\subsection{V-score}

\begin{figure}[tb]
\includegraphics[width=\linewidth]{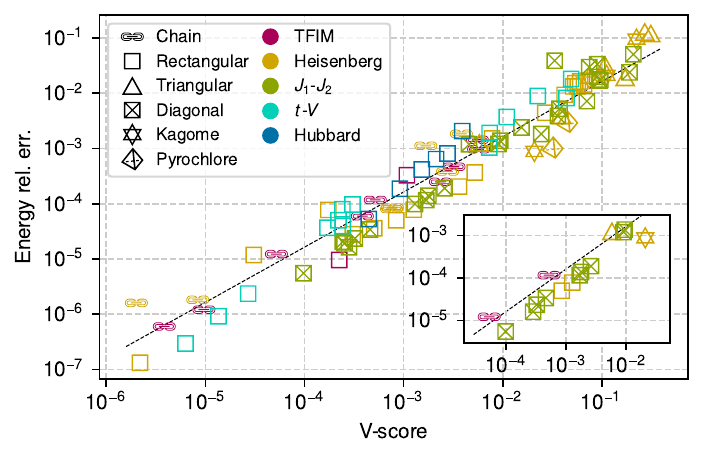}
\caption{\textbf{Validation of V-score against exact results.}
We compare V-scores versus energy relative errors on various strongly correlated models for which exact results (ED or QMC) are available.
The black dashed line is a least-squares fit of $\log(\text{energy rel. err.}) = \log(\text{V-score}) + C$, where $C = -1.80 \pm 0.08$.
The inset focuses on PQC results run on classical hardware (no shot noise included).
The symbol shapes and colors correspond to the lattice geometry and model type, respectively (see legend).
}
\label{fig:v-score-rel-err}
\end{figure}

\begin{figure*}[tb]
\hspace*{-0.7in} \includegraphics[width=0.95\linewidth]{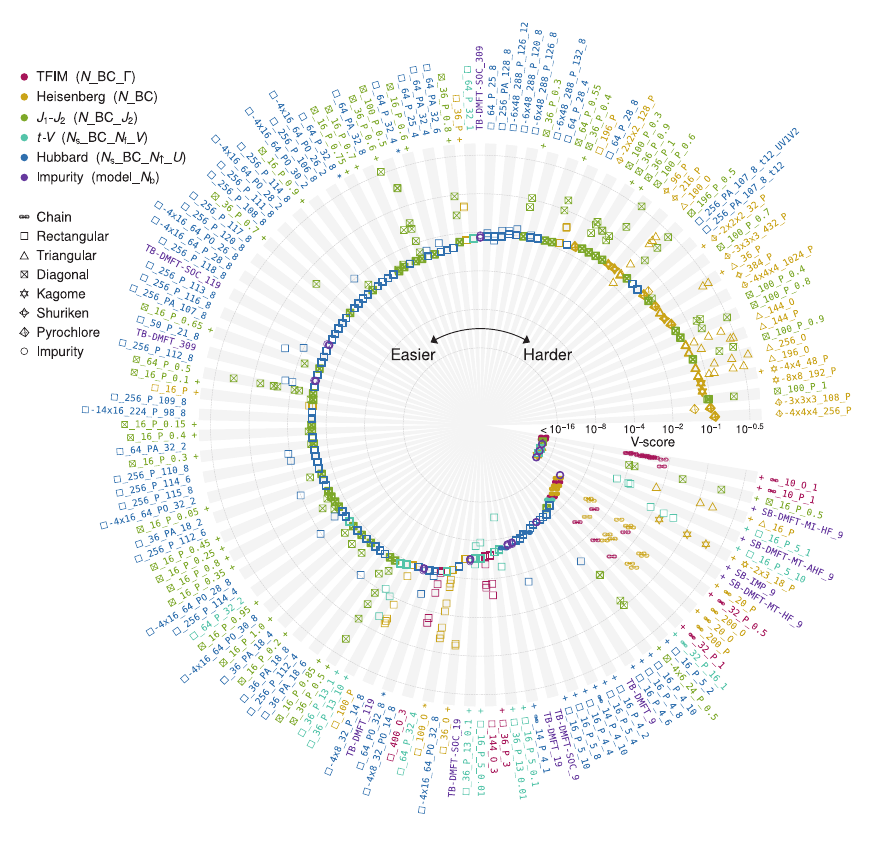}
\caption{\textbf{V-score rankings of the complete dataset of variational benchmark results.}
Each radial slice shows the results for a given Hamiltonian. The bold marker in the slice shows the method with the lowest variational energy (rather than the lowest V-score) for that Hamiltonian, and its V-score determines the clockwise ranking of that Hamiltonian.
The radial axis uses the doubly log scale to show the results across a wide range of V-score magnitude.
The results with V-scores less than $10^{-16}$ are indistinguishable from the exact solutions because of limited numerical precision.
The label outside each slice is the name of the Hamiltonian in the dataset, which describes the lattice geometry, the lattice size $N_\text{s}$, the boundary conditions (BCs), and the Hamiltonian parameters such as the interaction strength ($\Gamma$, $J_2$, $V$, $U$) and the number of particles ($N_\text{f}$, $N_\spinup$, $N_\text{b}$).
If a rectangular or diagonal lattice is non-square, its two edge lengths are shown such as ``\texttt{$\square$-4x8}''. Otherwise it is assumed to be square, and it can be tilted, allowing for numbers of spins such as $50$.
In a kagome or pyrochlore lattice, the number of copies of the unit cell in each spatial dimension is shown, such as ``\texttt{\ding{65}-4x4}''.
The BC can be O (open), P (periodic), or A (anti-periodic) on each dimension.
See legend for the meaning of the various parameters associated with each Hamiltonian type. Additionally, we include results on the generalized Hubbard model with second nearest neighbor hopping $t_2$, nearest neighbor repulse $V_1$, and second nearest neighbor repulse $V_2$, as indicated by the suffixes ``\texttt{\_t12}'' and ``\texttt{\_t12\_UV1V2}''. Formal definitions of the Hamiltonian parameters and specific types of impurity models are provided in the supplementary materials~\cite{smham}.
The ``\texttt{+}'' marker near the label indicates that an ED solution is available for that Hamiltonian, and the ``\texttt{\textasteriskcentered}'' marker indicates a numerically exact QMC solution.
}
\label{fig:v-score}
\end{figure*}

To quantify the accuracy of two or more variational methods applied on the same ground-state approximation task, a key indicator is the expectation value of the energy $E = \ev*{\hat{H}}$, an unbiased metric to assess the relative accuracy of variational methods; here, $\hat{H}$ is the relevant Hamiltonian. Given, for example, two independent methods preparing approximate ground states with variational energies $E_a$ and $E_b$, the one providing the lower energy can be considered more accurate.
From a practical point of view, however, it is preferable to have an absolute metric capable of predicting the accuracy of a method without comparing it with other methods. This would, for instance, allow comparing the performance of a given method on different tasks.
Nonetheless, it is unlikely we could find such a metric that is provably applicable in all cases, because its existence would also allow the solution of NP-hard problems~\cite{barahona1982computational}.
We are therefore forced to settle for an empirically applicable metric. Moreover, the metric should be easy to estimate with variational methods.

Apart from the mean energy, for most variational methods, we also have a controllable estimate of the energy variance $\Var E = \ev*{\hat{H}^2} - \ev*{\hat{H}}^2$.
It has the important property that it exactly vanishes if computed on the exact ground state. Therefore, $\Var E$ can be used to infer some information about the distance of the variational energy $E$ from the exact, and a priori unknown, ground-state energy $E_0$. After early empirical observations~\cite{kwon1993}, it has been shown that $\Var E$ scales linearly with the deviation $E - E_0$~\cite{imada2000,kashima2001,sorella2001}, so it can be used as a measure of the accuracy of the variational state.

We can use $E$ and $\Var E$ to create a dimensionless, intensive combination:
\begin{equation}
\text{V-score} := \frac{N \Var E}{(E - E_\infty)^2},
\end{equation}
where $N$ is the number of degrees of freedom, which is the number of spins for spin models, and the number of particles for fermionic models. The constant $E_\infty$ serves as a zero point of the energy, compensating for any global energy shift in the Hamiltonian definition.
The V-score is dimensionless in energy units and system size for the variational states that we consider and is also invariant under energy shifts by construction.
The procedure to benchmark the V-score is sketched in Fig.~\ref{fig:sketch}, and a detailed discussion of the definition of the V-score is presented in the supplementary materials~\cite{smvscore}.

\begin{figure}[tb]
\includegraphics[width=\linewidth]{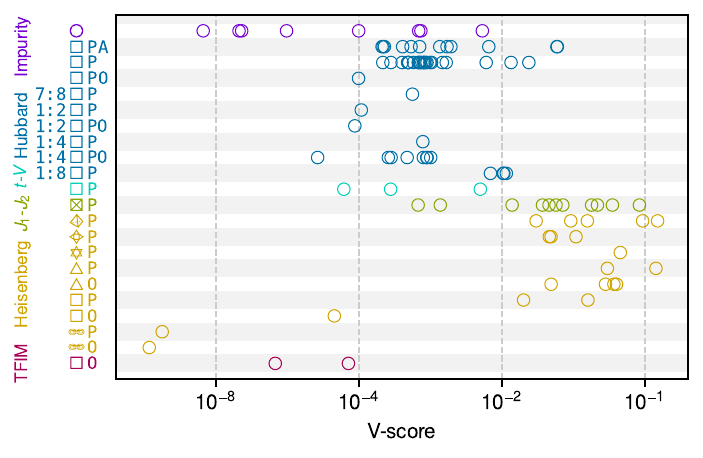}
\caption{\textbf{V-scores of Hamiltonians,} classified by Hamiltonian types and lattice geometries.
A Hamiltonian's V-score is defined as the V-score of the available variational method with the lowest energy on this Hamiltonian.
Only the Hamiltonians without ED results are shown.
The ratios to the left of the lattice icons are aspect ratios of rectangular lattices. The letters to the right are boundary conditions, which can be O (open), P (periodic), or A (anti-periodic) on each dimension.
}
\label{fig:v-score-ham}
\end{figure}

To further justify the definition of the V-score, in Fig.~\ref{fig:v-score-rel-err}, we present a comparison of this quantity against the energy relative error $(E - E_0) / (E_\infty - E_0)$ for a wide range of Hamiltonians and variational methods, where the ground-state energy $E_0$ is obtained by ED or numerically exact QMC.
Despite the great diversity of Hamiltonians and variational methods considered, the V-score is a notably consistent and reliable estimator for the order of magnitude of the energy relative error, as shown by the linear fit in Fig.~\ref{fig:v-score-rel-err}.
In the inset of Fig.~\ref{fig:v-score-rel-err}, we show that the same linear fit also well describes classically simulated PQCs, optimized with the variational quantum eigensolver (VQE) algorithm.
These results validate the V-score as an absolute performance metric for both classical and quantum variational algorithms, at least for the Hamiltonians and the techniques we consider in the paper.

\section{Identifying hard problems}

We can now discuss which Hamiltonians are hard for the state-of-the-art variational methods presented in our collection. Given the intrinsic exponential complexity of the problem, it is no longer possible to obtain ED results on larger system sizes. Thus, using the V-score as a guide in this task is crucial. In Fig.~\ref{fig:v-score} we show the V-score of all Hamiltonians and methods in our dataset.
We first select the best method for each Hamiltonian by choosing the one with the lowest variational energy. We then use this method's V-score as an absolute hardness metric in the ground-state approximation task, which we refer to as the V-score of this Hamiltonian. The V-score of the best-performing method is marked in bold in Fig.~\ref{fig:v-score}. For additional clarity, in Fig.~\ref{fig:v-score-ham}, we classify all those results by Hamiltonian types and lattice geometries.

It is well known that one-dimensional (1D) chain geometries are easy to solve with the density matrix renormalization group (DMRG). The small values of their V-scores in Fig.~\ref{fig:v-score} and Fig.~\ref{fig:v-score-ham} clearly label them as accessible, particularly for spin models.
Unfrustrated spin models typically also have small V-scores, ranging from $10^{-6}$ to $10^{-4}$ for the TFIM and the Heisenberg model on square lattices with open boundary conditions (OBCs). Moreover, these models can be efficiently simulated with unbiased stochastic techniques, such as QMC, and thus are easy to study on classical computers.
On the other hand, the V-scores show that frustrated geometries, such as triangular, kagome, pyrochlore, and the {\JoneJtwo} square lattice, as well as fermionic models, such as the Hubbard model, are the most demanding for variational algorithms. With the help of Fig.~\ref{fig:v-score-rel-err}, we can infer the order of magnitude of the energy relative error from the V-score. When this correspondence is applied to those hard problems, it predicts that we cannot expect an accuracy on the energy better than one or two digits.

\subsection{A perspective on quantum advantage}

Many recent theoretical and experimental efforts have been dedicated to showing the computational advantage of quantum computers over classical computers.
Informed by theoretical computer science arguments, random circuit sampling has been proposed as a specific task to show such advantage~\cite{boixo_characterizing_2018,arute_quantum_2019}. Still, it is unclear whether current noisy experimental quantum computing platforms can provide a solid and scalable advantage over classical ones~\cite{pan_solving_2022,gao_limitations_2021,aharonov_polynomial-time_2022}.
In addition to tasks of purely theoretical interest, there is also growing interest in finding practical quantum advantage~\cite{daley_practical_2022}, where quantum devices show a speedup for problems of scientific or technological relevance. The benchmarks introduced in this work belong to the family of approaches that can be useful to assess a quantum advantage that is practically relevant to physics.

In the context of variational ground-state algorithms, the V-score can readily be used as a good quality metric to assess quantum advantage.
Furthermore, the V-score can be used as an absolute indicator of hardness for Hamiltonians. In this respect, Hamiltonians with large classical V-scores are identified as hard problems that are not yet satisfactorily solved by classical computers and can be targeted by quantum computations.
Finally, in the absence of classical verifiability of the quantum solutions, the V-score can benchmark the progress of variational quantum computing-based approaches in solving ground-state problems relevant to physics.

In Fig.~\ref{fig:v-score}, we show the V-scores of the classical variational methods that we have analyzed.
From these results, we can infer that there is little room for quantum advantage in 1D geometries, where DMRG already achieves V-scores less than $10^{-8}$. In higher dimensions, unfrustrated spin models, such as the TFIM or the Heisenberg model on the square lattice, are similarly well approximated by existing classical methods, with V-scores less than $10^{-4}$.
On the contrary, specific regimes of higher-dimensional frustrated spin models constitute a clear challenge for existing classical methods. For example, the pyrochlore or kagome Heisenberg models typically present V-scores greater than $10^{-2}$---substantially higher than their unfrustrated counterparts. A similar scenario emerges for the Hubbard model in two dimensions with V-scores on the order of $10^{-2}$. In the specific regimes of interaction strengths, geometries, and frustration identified by large V-scores, these models represent natural targets for variational quantum algorithms. Impurity models with multiple bands also represent an ideal terrain for practical quantum advantage because the ability of classical algorithms to simulate them rapidly degrades upon increasing the number of bands, as Fig.~\ref{fig:v-score} shows for three-band models, and because of their importance for materials science.

We can also provide an early assessment of the V-scores obtained by the type of variational states that can be efficiently prepared on quantum computers. In this respect, it is encouraging to remark that PQCs perform well compared with classical variational methods, as shown in the inset of Fig.~\ref{fig:v-score-rel-err}, at least for the small system sizes that we consider, where PQCs can be classically simulated and ideally optimized. Applications on quantum hardware are much more challenging because of stochastic fluctuations and noise. However, the baseline of ideally optimized PQCs is promising for applications.

\section{Discussion and outlook}

In this work, we have introduced the V-score, an empirical metric to quantify the absolute accuracy of variational solutions to strongly interacting quantum models. Supplemented with state-of-the-art results obtained by a large variety of numerical methods, this metric allows us to clearly identify models, geometries, and regimes in which existing approaches are currently less accurate. With the introduction of future computational techniques and improved computing architectures, the outcomes of this analysis will naturally evolve in time, revealing in a certifiable manner the continuous improvements happening in the field. In this respect, the dataset presented in this work can be a standardized way of taking snapshots of the evolution of quantum many-body techniques with time.

Besides the importance of these benchmarks for future developments in computational techniques based on classical computers, it will be especially interesting to use the V-score to directly measure the impact of quantum computing-based approaches.
The hardest classical problem instances identified by large V-scores can be good candidates for studies based on quantum algorithms. In that context, the V-score can be used as a metric to assess progress in quantum variational state preparation, in the absence of classical verifiability.

\section*{Acknowledgments}

This paper is dedicated to the memory of Sandro Sorella, a dear friend, esteemed colleague, and co-author of this paper, who passed away before its submission.
We acknowledge discussions with A.~Sandvik and M.~Stoudenmire.
We thank L.~Reading-Ikkanda (Simons Foundation) for preparing Fig.~1.
The Flatiron Institute is a division of the Simons Foundation.

\subsection{Funding}

G.C.\ acknowledges support by the NCCR MARVEL, a National Centre of Competence in Research, funded by the Swiss National Science Foundation (grant no.~205602).
D.W.\ and G.C.\ acknowledge support from the Swiss National Science Foundation under grant no.~200336.
R.R.\ and G.C.\ acknowledge support from SEFRI through grant no.~MB22.00051 (NEQS - Neural Quantum Simulation).
N.A.\ and T.N.\ acknowledge support from the European Unions Horizon 2020 research and innovation program (ERC-StG-Neupert-757867-PARATOP) and from the Swiss National Science Foundation (grant PP00P2\_176877).
J.C.\ acknowledges support from the Natural Sciences and Engineering Research Council (NSERC), the Shared Hierarchical Academic Research Computing Network (SHARCNET), Compute Canada, and the Canadian Institute for Advanced Research (CIFAR) AI chair program.
M.I., Y.N., R.P., and M.S.\ acknowledge support by MEXT as ``Program for Promoting Researches on the Supercomputer Fugaku'' (Basic Science for Emergence and Functionality in Quantum Matter -- Innovative Strongly Correlated Electron Science by Integration of Fugaku and Frontier Experiments, JPMXP1020200104) together with computational resources of supercomputer Fugaku provided by the RIKEN Center for Computational Science (project ID: hp210163 and hp220166).
M.I.\ acknowledges the support of MEXT KAKENHI, Grant-in-Aid for Transformative Research Areas, grant nos.~JP22H05111 and JP22H05114 and JSPS KAKENHI grant no.~JP19H00658.
M.I.\ and Y.N.\ acknowledge support from MEXT as ``Program for Promoting Researches on the Supercomputer Fugaku'' (grant no.~JPMXP1020230411).
G.M.\ acknowledges support from the Swiss National Science Foundation (grant PCEFP2\_203455).
Y.N.\ acknowledges support from Grant-in-Aids for Scientific Research (JSPS KAKENHI) (grant nos.~JP23H04869, JP23H04519, JP23K03307, and JP21H01041) and JST (grant no.~JPMJPF2221).
J.M.S.\ and S.R.W.\ acknowledge the support of the US National Science Foundation under grant DMR-2110041.
L.W.\ and Q.Y.\ acknowledge the support of the National Natural Science Foundation of China under grant nos.~92270107, 12188101, T2225018, and T2121001.
A.W.\ acknowledges support by the DFG through the Emmy Noether programme (509755282).
Y.Y.\ thanks the Center for Computational Quantum Physics (CCQ) of the Flatiron Institute, Simons Foundation for support and hospitality.
Research at the Perimeter Institute is supported in part by the government of Canada through the Department of Innovation, Science and Economic Development and by the Province of Ontario through the Ministry of Colleges and Universities.

\subsection{Author contributions}

D.W.\ implemented the research infrastructure and curated the data.
R.R.\ designed the final form of the V-score.
G.C.\ designed and coordinated the project, with the help of M.I.\ and R.R.
The main text was written by D.W., R.R., F.V., and G.C.\ with input from all authors. All authors contributed to the supplementary materials.
A detailed record of the authors' contributions to the dataset can be found on the GitHub repository.

\subsection{Competing interests}

The authors declare no competing interests.

\subsection{Data and materials availability}

The VarBench dataset and the code to analyze the benchmark results and reproduce the figures in this manuscript are available at Zenodo~\cite{varbench}.
Every benchmark result either has a reference to a paper reporting it or can be reproduced using the code at Zenodo~\cite{varbench_methods}.
The software packages used to reproduce the data include ITensor~\cite{itensor}, mVMC~\cite{MISAWA2019447}, and NetKet~\cite{netket3:2021}. Future updates to the dataset can be found at \texttt{https://github.com/varbench/varbench} .

\clearpage

\onecolumngrid
\begin{center}
\textbf{\large Supplementary Materials for ``Variational Benchmarks for \\ Quantum Many-Body Problems''}
\end{center}
\vspace{4ex}
\twocolumngrid

\setcounter{section}{0}
\setcounter{equation}{0}
\setcounter{figure}{0}

\renewcommand{\thesection}{S\arabic{section}}
\renewcommand{\theequation}{S\arabic{equation}}
\renewcommand{\thefigure}{S\arabic{figure}}

\section{Overview of the many-body Hamiltonians}

Here we provide the definitions of the many-body quantum Hamiltonians used for benchmarking purposes.

\subsection{Spin models}

The transverse-field Ising model (TFIM) is
\begin{equation}
\hat{H} = J \sum_{\langle i, j \rangle} \hat{\sigma}^z_i \hat{\sigma}^z_j
+ \Gamma \sum_i \hat{\sigma}^x_i,
\label{eq:tfim}
\end{equation}
where $\langle i, j \rangle$ runs over nearest neighbors, $\Gamma$ is the transverse field strength, and $\hat{\sigma}^x, \hat{\sigma}^y, \hat{\sigma}^z$ are Pauli matrices. In the dataset we use $J = 1$.

The Heisenberg model is
\begin{equation}
\hat{H} = J \sum_{\langle i, j \rangle, a} \hat{\sigma}^a_i \hat{\sigma}^a_j,
\end{equation}
where $a$ runs in $\{x, y, z\}$.

The {\JoneJtwo} model is
\begin{equation}
\hat{H} = J_1 \sum_{\langle i, j \rangle, a} \hat{\sigma}^a_i \hat{\sigma}^a_j
+ J_2 \sum_{\llangle i, j \rrangle, a} \hat{\sigma}^a_i \hat{\sigma}^a_j,
\end{equation}
where $\llangle i, j \rrangle$ runs over next-nearest neighbors (diagonals if the lattice is 2D square), and $J_2$ is the next-nearest neighbor interaction. In the dataset we use $J_1 = 1$.

\subsection{Fermions}

The $t$-$V$ model is
\begin{equation}
\hat{H} = -t \sum_{\langle i, j \rangle} \left( \hat{c}^\dagger_i \hat{c}_j + \hat{c}^\dagger_j \hat{c}_i \right)
+ V \sum_{\langle i, j \rangle} \hat{n}_i \hat{n}_j,
\end{equation}
where $V$ is the Coulomb repulsive interaction strength, and the number of fermions is fixed to $N_\text{f}$. In the dataset we consider $t = 1$.

The Hubbard model is
\begin{equation}
\hat{H} = -t \sum_{\langle i, j \rangle, \sigma} \left( \hat{c}^\dagger_{i \sigma} \hat{c}_{j \sigma} + \hat{c}^\dagger_{j \sigma} \hat{c}_{i \sigma} \right)
+ U \sum_i \hat{n}_{i \spinup} \hat{n}_{i \spindown},
\end{equation}
where $U$ is the on-site interaction strength, and the numbers of fermions are fixed to $N_\spinup$ and $N_\spindown$. We only consider the case of $N_\spinup = N_\spindown$.

Additionally, we include the generalized Hubbard model with second nearest neighbor hopping $t_2$ and repulse $V_2$:
\begin{align}
\hat{H} = &-t_1 \sum_{\langle i, j \rangle, \sigma} \left( \hat{c}^\dagger_{i \sigma} \hat{c}_{j \sigma} + \hat{c}^\dagger_{j \sigma} \hat{c}_{i \sigma} \right)
- t_2 \sum_{\llangle i, j \rrangle, \sigma} \left( \hat{c}^\dagger_{i \sigma} \hat{c}_{j \sigma} + \hat{c}^\dagger_{j \sigma} \hat{c}_{i \sigma} \right) \nonumber \\
&+U \sum_i \hat{n}_{i \spinup} \hat{n}_{i \spindown}
+ V_1 \sum_{\langle i, j \rangle, \sigma} \hat{n}_{i \sigma} \hat{n}_{j \sigma}
+ V_2 \sum_{\llangle i, j \rrangle, \sigma} \hat{n}_{i \sigma} \hat{n}_{j \sigma}.
\end{align}
We use $t_1 = 1$, $t_2 = -0.25$, $V_1 = 1$, $V_2 = 0.5$.

\subsection{Impurity models} \label{sec:impurity-models-def}

A typical Anderson impurity Hamiltonian $\hat{H}_\text{A}$ contains two parts
\begin{align}
\hat{H}_\text{A} &:= \hat{H}_\text{loc} + \hat{H}_\text{bath}, \\
\hat{H}_\text{loc} &:= \phantom{+{}}\sum_{\{\alpha\}} \epsilon^0_{\alpha_1 \alpha_2} \hat{d}^\dagger_{\alpha_1} \hat{d}_{\alpha_2} \nonumber \\
&\phantom{:={}}+ \sum_{\{\alpha\}} U_{\alpha_1 \alpha_2 \alpha_3 \alpha_4} \hat{d}^\dagger_{\alpha_1} \hat{d}^\dagger_{\alpha_2} \hat{d}_{\alpha_3} \hat{d}_{\alpha_4}, \\
\hat{H}_\text{bath} &:= \phantom{+{}}\sum_{\{\alpha\}} \sum_{l = 1}^{N_\text{b}} \epsilon^l_{\alpha_1 \alpha_2} \hat{c}^\dagger_{l \alpha_1} \hat{c}_{l \alpha_2} \nonumber \\
&\phantom{:={}}+ \sum_{\{\alpha\}} \sum_{l = 1}^{N_\text{b}} \left( \nu_{\alpha_1 \alpha_2}^l \hat{d}^\dagger_{\alpha_1} \hat{c}_{l \alpha_2} + \text{h.c.} \right),
\label{eq:imp}
\end{align}
where a locally interacting impurity $\hat{H}_\text{loc}$ is coupled to a non-interacting bath $\hat{H}_\text{bath}$. The indices $\{\alpha\}$ are a collection of quantum numbers denoting the impurity (or the $l$-th bath site) degrees of freedom of the fermionic creation $\hat{d}^\dagger$ (or $\hat{c}_l^\dagger$) and annihilation $\hat{d}$ (or $\hat{c}_l$) operators, and $N_\text{b}$ is the number of bath sites per spin-orbital. The bath parameters $\{\epsilon^l, \nu^l\}$ are connected to the hybridization function as $\Delta(\omega) = \sum_{l = 1}^{N_\text{b}} \frac{\nu^l \nu^{l \dagger}}{\omega - \epsilon^l}$, and can be obtained by discretizing the hybridization function $\Delta(\omega)$ on the real frequency axis into $N_\text{b}$ equidistant intervals $\{I_l\}$ of size $\Delta \omega$ as
\begin{align}
\epsilon^l_{\alpha \alpha} &= \min I_l + \frac{\Delta \omega}{2}, \\
|\nu^l|^2 &= \int_{I_l} \dd \omega \left( -\frac{\Im \Delta(\omega)}{\pi} \right).
\end{align}

We consider two types of interactions that are frequently encountered in DMFT calculations: the single-band Hubbard interaction
\begin{equation}
\hat{H}_\text{loc} = U \hat{n}_\spinup \hat{n}_\spindown + \epsilon^0 \left( \hat{n}_\spinup + \hat{n}_\spindown \right),
\end{equation}
where $\hat{n}_\sigma = \hat{d}^\dagger_\sigma \hat{d}_\sigma$ is the particle number operator, with $\sigma \in \left\{\spinup, \spindown\right\}$; the three-band rotationally invariant Kanamori interaction~\cite{georges2013strong}
\begin{align}
\hat{H}_\text{loc} &\phantom{:}= \hat{H}_\text{DD} + \hat{H}_\text{SF} + \hat{H}_\text{PH}, \\
\hat{H}_\text{DD} &:= \phantom{+{}}U \sum_m \hat{n}_{m \spinup} \hat{n}_{m \spindown} \nonumber \\
&\phantom{:={}}+ (U - 2 J) \sum_{m' > m, \sigma} \hat{n}_{m \sigma} \hat{n}_{m' \bar{\sigma}} \nonumber \\
&\phantom{:={}}+ (U - 3 J) \sum_{m' > m, \sigma} \hat{n}_{m \sigma} \hat{n}_{m' \sigma}, \\
\hat{H}_\text{SF} &:= J \sum_{m' m} \left( \hat{d}^\dagger_{m \spinup} \hat{d}_{m \spindown} \hat{d}_{m' \spinup} \hat{d}^\dagger_{m' \spindown} + \text{h.c.} \right), \\
\hat{H}_\text{PH} &:= -J \sum_{m' > m} \left( \hat{d}^\dagger_{m \spinup} \hat{d}^\dagger_{m \spindown} \hat{d}_{m' \spinup} \hat{d}_{m' \spindown} + \text{h.c.} \right),
\label{eq:kanamori}
\end{align}
with $m \in \{1, 2, 3\}$ being the orbital index, and $\hat{H}_\text{DD}$, $\hat{H}_\text{SF}$, and $\hat{H}_\text{PH}$ denoting the density-density, the spin-flip, and the pair-hopping interactions respectively.

The following models representing a collection of typical solutions in practical DMFT calculations are considered:
(\textbf{SB-Imp}) single-band Anderson impurity model with a semielliptic spectral function, i.e., $-\frac{1}{\pi} \Im \Delta(\omega) = \frac{2}{\pi D} \sqrt{1 - \left( \frac{\omega}{D} \right)^2}$ with $D$ being the half-bandwidth and $U = D$;
(\textbf{SB-DMFT-MT-HF}) DMFT metal solution of the single band Hubbard model on the Bethe lattice with $U = 2 D$ at half-filling $n = 1$ and
(\textbf{SB-DMFT-MT-AHF}) doped case $n = 0.8$;
(\textbf{SB-DMFT-MI-HF}) DMFT Mott-insulator solution of the single band Hubbard model on the Bethe lattice with $U = 4 D$ at half-filling $n = 1$;
three-band models with Kanamori interaction $U = 2.3\ \text{eV}$ and $J = 0.4\ \text{eV}$ that are based on the material-realistic DMFT solutions of the archetypal Hund's metal Sr$_2$RuO$_4$ in the $t_{2 g}$ subspace (\textbf{TB-DMFT-SOC}) with and (\textbf{TB-DMFT}) without spin-orbit coupling.

\section{V-score}

\subsection{Definition and justification of the V-score for lattice models} \label{sec:v-score-def}

Given the observed energy expectation $E = \ev{\hat{H}}{\psi}$ and variance $\Var E = \ev{\hat{H}^2}{\psi} - \ev{\hat{H}}{\psi}^2$ of a variational quantum state $\ket{\psi}$, we want to introduce a function of these two quantities, the V-score, as a metric quantifying how close $E$ is to the ground-state energy. As the exact ground-state energy is unknown in general, the mean energy alone is not enough to characterize the quality of a variational optimization. The energy variance is zero for an eigenstate of the Hamiltonian, thus, also for the ground state. Therefore, assuming that the variational optimization does not converge towards an excited state, we can infer that the V-score should be a monotonic function of $\Var E$.

To start with, we investigate how these two observables scale asymptotically with the number of degrees of freedom $N$. For any well-defined variational state, $E$ scales linearly with $N$, as the energy is an extensive thermodynamic quantity.
To analyze the scaling of $\Var E$, we evoke the cluster property of the variational state. The Hamiltonian $\hat{H}$ is written as a sum of $N_H = O(N)$ local terms, $\hat{H} = \sum_{i = 1}^{N_H} \hat{h}_i$. If the correlations of these local terms satisfy the cluster property
\begin{equation}
\left| \ev*{\hat{h}_i \hat{h}_j} - \ev*{\hat{h}_i} \ev*{\hat{h}_j} \right| \le \frac{A}{d(i, j)^{D + \epsilon}},
\end{equation}
where $A, \epsilon > 0$, $d$ is a distance function on the lattice, and $D$ is the space dimension, then $\Var E$ scales linearly with $N$. We can therefore construct a dimensionless number,
\begin{equation}
\frac{N \Var E}{E^2},
\label{eq:v-score-1}
\end{equation}
that does not scale with the energy unit or with $N$ asymptotically. Moreover, for variational optimizations converging towards the ground state, $\Var E$ can be expected to scale linearly with the energy difference from the ground state~\cite{imada2000,kashima2001,sorella2001,PhysRevC.67.041301,taddei2015iterative}. Therefore, the dimensionless number in \Eq{v-score-1} linearly quantifies the energy difference as well.

However, the expression in \Eq{v-score-1} is still prone to a shift of energy $\hat{H} \mapsto \hat{H} + C$, where $C$ is an arbitrary constant. In particular, if we fine-tune $C$ such that the ground-state energy is zero, \Eq{v-score-1} can be expected to scale \emph{inversely} with the energy difference from the ground state. Similarly, if $C$ is such that $E = 0$, \Eq{v-score-1} is ill-defined. To solve this issue, we need to fix a zero point of energy $E_\infty$ in the definition of the V-score:
\begin{equation}
\text{V-score} := \frac{N \Var E}{(E - E_\infty)^2},
\label{eq:v-score}
\end{equation}
such that the average energy $E$ of a variational ground-state wave function $\ket{\psi}$ is always different from $E_\infty$.
In this work we choose $E_\infty$ to be the energy expectation of a random state (sampled uniformly on the unit sphere surface) in the Hilbert (sub)space $\mathcal{H}$, because an optimized variational state typically has lower energy than a random state. $E_\infty$ is also the energy expectation of a thermal state at infinite temperature restricted to $\mathcal{H}$, which can be computed from the trace of $\hat{H}$:
\begin{equation}
E_\infty := \frac{\Tr \hat{H}}{\dim \mathcal{H}},
\label{eq:E-infty-trace}
\end{equation}
where $\dim \mathcal{H}$ is the dimension of the Hilbert (sub)space. For the models we consider in this work, the dimension of the Hilbert space is finite for a finite lattice size, and $E_\infty$ is a finite number.

We now discuss our choices for the number of degrees of freedom $N$. For unconstrained spin-$1/2$ Hilbert spaces, we define $N$ to be equal to the number of lattice sites $N_\text{s}$.
For the $t$-$V$ model with fixed particle number, $N$ is defined to be equal to the particle number $N_\text{f}$, and we have $\dim \mathcal{H} = B(N_\text{s}, N_\text{f})$, where $B$ is the binomial coefficient.
For the Hubbard model with fixed particle numbers, $N$ is defined to be equal to the sum of the numbers of spin up and down fermions, $N = N_\spinup + N_\spindown$, while $\dim \mathcal{H} = B(N_\text{s}, N_\spinup)\,B(N_\text{s}, N_\spindown)$.
We remark that the V-score can also be applied to estimations of the lowest-energy excited states in symmetry sectors different from the ground state symmetry sector.

This choice of energy shift $E_\infty$ supposes that the model contains only relevant low-energy degrees of freedom. Actually, if we add many high-energy states, they will have the effect of artificially raising $E_\infty$ without contributing much to the ground state. Therefore, for some models we do not consider in this work (e.g.\ bosonic models or quantum chemical models), a cutoff on the relevant energy scale must be set into place in order to use this definition of $E_\infty$. An alternative strategy to define $E_\infty$ for these models would be to use the mean-field energy, which is not affected by the problem of having high-energy states. However, it is generally not straightforward to compute the mean-field energy as it is in itself an NP-hard problem~\cite{schuch2009computational}, there exist many variants of mean-field theory, and it could make weak-coupling calculations beyond mean-field theory artificially hard. Impurity models, which we have introduced in Sec.~\ref{sec:impurity-models-def}, require an adapted definition of $E_\infty$, see Sec.~\ref{sec:E-infty-impurity}.

\subsection{Calculation of $E_\infty$ for lattice models}

\subsubsection{Analytical formulae for specific models}

For quantum spin models, we have $E_\infty = 0$ when the Hamiltonian is written as a sum of Pauli strings with no term proportional to the identity operator, as all Pauli matrices are traceless.
For the spinless $t$-$V$ model with fixed particle number, only the diagonal term $V \sum_{\langle i, j \rangle} \hat{n}_i \hat{n}_j$ contributes, and we have
\begin{equation}
E_\infty = \frac{V |\mathcal{E}| N_\text{f} (N_\text{f} - 1)}{N_\text{s} (N_\text{s} - 1)},
\end{equation}
where $|\mathcal{E}|$ is the number of nearest neighbor bonds.
For the Hubbard model with fixed particle numbers, only the diagonal term $U \sum_i \hat{n}_{i \spinup} \hat{n}_{i \spindown}$ contributes, and we have
\begin{equation}
E_\infty = \frac{U N_\spinup N_\spindown}{N_\text{s}}.
\end{equation}
Apart from fixing the number of fermions, in this work we do not consider symmetries of the Hamiltonians when calculating $E_\infty$.

\subsubsection{General case} \label{sec:E-infty-sampling}

It is generally efficient to get a numerical estimate of $E_\infty$ with stochastic methods. For simplicity, we limit our discussion to spinless fermions and a short-range translation-invariant Hamiltonian $\hat{H}$. Considering a Hilbert space $\mathcal{H}$ with fixed particle number $N_\text{f}$ and lattice size $N_\text{s}$, we estimate $E_\infty$ by sampling uniformly a bit string $x = (x_1, \ldots, x_{N_\text{s}})$ with $x_i \in \{0, 1\}$ and the constraint $\sum_i x_i = N_\text{f}$, then taking the average
\begin{equation}
E_\infty = \bigl\langle \ev{\hat{H}}{x} \bigr\rangle_{x \sim \mathcal{U}_{N_\text{s}, N_\text{f}}(x)},
\label{eq:E-infty-sampling}
\end{equation}
where $\ket{x}$ is an element of the Fock basis and $\mathcal{U}_{N_\text{s}, N_\text{f}}(x)$ is the aforementioned uniform distribution.
The variance of the estimator of $E_\infty$ can be written as
\begin{equation}
\Var E_\infty = \bigl\langle \ev{\hat{H}}{x}^2 \bigr\rangle_{x \sim \mathcal{U}_{N_\text{s}, N_\text{f}}(x)} - E_\infty^2.
\end{equation}
Comparing to the physical variance of the energy at infinite temperature
\begin{equation}
\Var_\mathcal{H} \hat{H} := \frac{\Tr \hat{H}^2}{\dim \mathcal{H}} - E_\infty^2,
\end{equation}
which can be computed by the same sampling method:
\begin{equation}
\Var_\mathcal{H} \hat{H} = \bigl\langle \ev{\hat{H^2}}{x} \bigr\rangle_{x \sim \mathcal{U}_{N_\text{s}, N_\text{f}}(x)} - E_\infty^2,
\end{equation}
we have $\Var E_\infty \le \Var_\mathcal{H} \hat{H}$.
As $\hat{H}$ is short-range and translation invariant, and as $\ket{x}$ satisfies the cluster property being a product state, we have the scalings
\begin{equation}
E_\infty = O(N), \quad \Var_\mathcal{H} \hat{H} = O(N),
\end{equation}
which implies that the one-sample stochastic relative error on $E_\infty$ vanishes with increasing system size:
\begin{equation}
\frac{\sqrt{\Var E_\infty}}{|E_\infty|} \le \frac{\sqrt{\Var_\mathcal{H} \hat{H}}}{|E_\infty|} = O\left(\frac{1}{\sqrt{N}}\right).
\end{equation}
This shows that, in the thermodynamic limit $N \to \infty$, even just one sample is enough to estimate $E_\infty$. Moreover, the calculation of $\ev{\hat{H}}{x}$ can be done in a computational time increasing linearly with the number of lattice sites. Therefore, we conclude that the statistical procedure we discussed is efficient.

\subsection{Bounds on the V-score}

We now consider bounds on the ratio of the V-score and the energy relative error. The lower bound is obtained when the variational state exactly coincides with an excited state, which is therefore zero. In order to prove an upper bound, we maximize $\Var E$ given a fixed mean energy $E$. When the spectrum is bounded from above, e.g., for finite systems, $\Var E$ is maximized, at fixed average energy $E$, when the variational state $\ket{\psi}$ is a linear combination of the ground state and the maximal energy state
\begin{equation}
\ket{\psi} = \sqrt{1 - I} \ket{E_0} + \sqrt{I} \ket{E_M},
\end{equation}
where $E_0, \ket{E_0}$ and $E_M, \ket{E_M}$ are the minimal and the maximal eigenvalue-eigenvector pairs respectively, $I \in [0, 1]$ is an interpolation parameter such that $E = (1 - I) E_0 + I E_M$, and we have assumed that both $\hat{H}$ and $-\hat{H}$ are non-degenerate for simplicity of notation. The energy variance of $\ket{\psi}$ is equal to $(E_M - E) (E - E_0)$. When we express the variance in terms of the V-score, we reach the following bound for the ratio of the V-score and the relative energy error:
\begin{equation}
0 \le \frac{\text{V-score}}{(E - E_0) / (E_\infty - E_0)} \le N \frac{(E_\infty - E_0) (E_M - E)}{(E_\infty - E)^2}.
\label{eq:v-score-bound}
\end{equation}
\Eq{v-score-bound} shows that, while the V-score can arbitrarily underestimate the difference from the ground-state energy, it cannot arbitrarily overestimate it. We remark that the upper bound is linear in system size, as the aforementioned variational state does not respect the cluster property for energy correlations.

\subsection{Scaling of the V-score in the limit of vanishing ground state infidelity}

In this section, we adapt the argument from Ref.~\cite{taddei2015iterative} to show the linear scaling of the V-score with the energy relative error in the limit where the ground state infidelity goes to zero. We remark that this result is not directly applicable to large system sizes as the ground state infidelity is expected to grow exponentially with system size for a fixed energy relative error.

We consider a variational state parameterized by a control parameter $c \ge 0$:
\begin{equation}
\ket{\psi(c)} := \sum_i \psi_i(c) \ket{E_i},
\end{equation}
where $\psi_i(c) \in \mathbb{C}$ and $\ket{E_i}$ are energy eigenstates with eigenvalue $E_i$. We suppose $E_0 \le E_i$, and we define the set of ground state indices as $G := \{i \mid E_i = E_0\}$. We introduce the ground state infidelity
\begin{equation}
I(c) := \frac{\sum_{i \notin G} |\psi_i(c)|^2}{\sum_i |\psi_i(c)|^2}.
\end{equation}
We suppose that when the control parameter $c$ goes to infinity, the ground state infidelity goes to zero:
\begin{equation}
\lim_{c \to \infty} I(c) = 0.
\end{equation}
We also suppose that $I(c) \neq 0$ for every $c$.

In the following, we determine the scaling of the V-score when $c \to \infty$.
For an operator $\hat{O}$, we define
\begin{align}
\ev*{\hat{O}}_c &:= \frac{\ev{\hat{O}}{\psi(c)}}{\ip{\psi(c)}}, \\
\ev*{\hat{O}}_{\text{ex}, c} &:= \frac{\sum_{i \notin G} |\psi_i(c)|^2 \ev{\hat{O}}{E_i}}{\sum_{i \notin G} |\psi_i(c)|^2}.
\end{align}
Then we have
\begin{equation}
\ev*{\hat{H}}_c - E_0 = (\ev*{\hat{H}}_{\text{ex}, c} - E_0) I(c).
\end{equation}
We also define
\begin{align}
\sigma_c^2 :={}&\ev*{\hat{H}^2}_c - \ev*{\hat{H}}_c^2 \\
={}&\ev*{(\hat{H} - E_0)^2}_c - (\ev*{\hat{H}}_c - E_0)^2 \\
={}&\ev*{(\hat{H} - E_0)^2}_{\text{ex}, c}\,I(c) - \bigl( (\ev*{\hat{H}}_{\text{ex}, c} - E_0) I(c) \bigr)^2 \\
={}&\ev*{(\hat{H} - E_0)^2}_{\text{ex}, c}\,I(c) \Bigl( 1 + O\bigl( I(c) \bigr) \Bigr),
\end{align}
where we used $0 < (\ev*{\hat{H}}_{\text{ex}, c} - E_0)^2 \le \ev*{(\hat{H} - E_0)^2}_{\text{ex}, c}$. We then write
\begin{align}
&\lim_{c \to \infty} \frac{\text{V-score}}{(\ev*{\hat{H}}_c - E_0) / (E_\infty - E_0)} \nonumber \\
= &\lim_{c \to \infty} \frac{N \sigma_c^2 (E_\infty - E_0)}{(\ev*{\hat{H}}_c - E_0) (E_\infty - \ev*{\hat{H}}_c)^2} \\
= &\lim_{c \to \infty} \frac{N \ev*{(\hat{H} - E_0)^2}_{\text{ex}, c} (E_\infty - E_0)}{(\ev*{\hat{H}}_{\text{ex}, c} - E_0) (E_\infty - \ev*{\hat{H}}_c)^2} \\
= &\lim_{c \to \infty} \frac{N \ev*{(\hat{H} - E_0)^2}_{\text{ex}, c}}{(\ev*{\hat{H}}_{\text{ex}, c} - E_0) (E_\infty - E_0)}.
\end{align}
If we suppose that the following limits exist:
\begin{align}
\sigma^2_\text{ex} &:= \lim_{c \to \infty} \bigl( \ev*{\hat{H}^2}_{\text{ex}, c} - \ev*{\hat{H}}_{\text{ex}, c}^2 \bigr), \\
E_\text{ex} &:= \lim_{c \to \infty} \ev*{\hat{H}}_{\text{ex}, c},
\end{align}
then we have
\begin{equation}
\lim_{c \to \infty} \frac{\text{V-score}}{(\ev*{\hat{H}}_c - E_0) / (E_\infty - E_0)}
= \frac{N \bigl( \sigma^2_\text{ex} + (E_\text{ex} - E_0)^2 \bigr)}{(E_\text{ex} - E_0) (E_\infty - E_0)}.
\end{equation}
This limit is non-singular because
\begin{gather}
E_0 + \Delta \le E_\text{ex} \le E_M, \\
0 \le \sigma^2_\text{ex} \le \frac{1}{4} (E_M - E_0 - \Delta)^2,
\end{gather}
where $\Delta$ is the gap between the ground state subspace and the rest of the spectrum, and $E_M$ is the maximal energy. If we suppose further that $E_\text{ex} = E_0 + \Delta$ and $\sigma_\text{ex}^2 = 0$, which is valid, e.g., for two-level systems or for a thermal state at temperature $T=1/c$, then we have
\begin{equation}
\lim_{c \to \infty} \frac{\text{V-score}}{(\ev*{\hat{H}}_c - E_0) / (E_\infty - E_0)} = \frac{N \Delta}{E_\infty - E_0}.
\end{equation}

\subsection{$E_\infty$ for impurity models} \label{sec:E-infty-impurity}

As mentioned in Sec.~\ref{sec:v-score-def}, when applied to impurity models, the definition of $E_\infty$ should be modified such that $\ket{x}$ in \Eq{E-infty-sampling} is restricted to the low-energy subspace. For impurity models, there is a natural way to filter out high-energy states that do not contribute to the ground state. Noticing that the fast convergence of density of bath sites with negative (positive) on-site energy to be fully occupied (empty), $\ket{x}$ can be restricted to the low-energy subspace as $\ket{x} = \ket{\text{o}} \otimes \ket{\text{e}} \otimes \ket{i}$, where $\ket{\text{o}}$ ($\ket{\text{e}}$) denotes the product state of completely occupied (empty) bath sites, $\ket{i} \in \mathcal{H}_\text{imp} \otimes \mathcal{H}_\text{a}$ is a product state belonging to the Hilbert space composed by the impurity and the active bath site. The active bath site is determined by the one with the smallest absolute on-site energy. Following \Eq{E-infty-trace}, $E_\infty$ is modified for impurity models as
\begin{equation}
E_\infty = \frac{\Tr \hat{\rho}_\text{b} \otimes \hat{\rho}_i \hat{H}_\text{A}}{\dim \mathcal{H}_\text{imp}},
\end{equation}
with $\hat{\rho}_\text{b} = \ket{\text{e}}\ket{\text{o}}\bra{\text{o}}\bra{\text{e}}$ and $\hat{\rho}_i = \ket{i}\bra{i}$. For an impurity model with $N_\text{e}$ electrons and $N_\text{o}$ occupied bath sites, we evaluate $E_\infty$ using the stochastic sampling method presented in Sec.~\ref{sec:E-infty-sampling} with the constraint that each sampled state $\ket{i}$ has a fixed particle number of $N_\text{b} - N_\text{o}$.

\section{Overview of the numerical methods}

\subsection{Exact diagonalization}

The quantum many-body problem can be solved numerically to arbitrary precision on small lattice sizes using exact diagonalization without any approximation. Typically, an iterative algorithm like the Lanczos algorithm~\cite{Lanczos1950} is used to solve for the eigenvalues and eigenvectors of the static Schr\"odinger equation
\begin{equation}
\hat{H} \ket{\psi} = E \ket{\psi}.
\end{equation}
The Lanczos algorithm has proven to be a reliable tool for computing ground-state energies up to machine precision. Hence, data retrieved from exact diagonalization can be considered exact. To achieve the currently largest system sizes, both memory and CPU time limitations must be dealt with. To avoid memory bottlenecks, the matrix-vector multiplication operations in the Lanczos algorithm are performed ``on-the-fly'', i.e., without storing the Hamiltonian matrix, neither in full nor in sparse format, but by implementing a matrix-vector multiplication function. Moreover, using Hamiltonian symmetries allows block-diagonalization to reduce the memory footprint further. Using a symmetry-adapted basis requires efficient algorithms to evaluate matrix elements in this basis, e.g.\ sublattice coding algorithms. Finally, also large-scale parallelization for distributed memory computers is required, which poses challenges in managing load balancing. The necessary techniques required to achieve the currently largest system sizes are explained in detail in Ref.~\cite{Wietek2018}.

\subsection{Tensor networks}

\subsubsection{Matrix product states and density matrix renormalization group}

The density matrix renormalization group (DMRG) is a variational technique, first introduced by White in 1992 to accurately describe the ground-state properties of one-dimensional (1D) quantum lattices~\cite{white_density_1992}. While exact diagonalization methods operate in an exponentially large basis, DMRG works with the degrees of freedom tied to a few sites at a time. At the heart of the DMRG algorithm is the matrix product state (MPS) ansatz~\cite{DMRGinAGEofMPS}, which represents the many-indexed wave function as a chain of tensors, one for each site, with links connecting the sites in a 1D layout. DMRG is an extremely efficient procedure for optimizing the coefficients of the MPS. The required \emph{bond dimension}, i.e.\ dimension of the indices linking the tensors, is determined by the degree of entanglement in the state being described. In the calculations presented here, which do not utilize parallelization beyond a single node, bond dimensions are limited to about $10,000$. The required bond dimension for the ground state of a model system is thus tied to the area law of entanglement, which states that the entanglement entropy of a bipartition of a system varies as the size of the boundary rather than the volume of either subsystem~\cite{AreaLaw}. DMRG is ideal for gapped 1D systems, where the area law implies that the bond dimension (for a fixed error) is independent of the length of the chain. For a gapless chain, there is a slow logarithmic growth of the bond dimension with length; nevertheless, spin-$1/2$ chains with lengths in the thousands are still rather easy on a laptop.

For two-dimensional (2D) clusters, the area law implies that the bond dimension is independent of the length but grows exponentially with the width. Despite the exponential, the general efficiency and robustness of DMRG makes it one of the most powerful and versatile methods for studying many 2D lattice models~\cite{2DDMRG}. Success has required developing a variety of specific techniques and ``tricks''; for example, the standard two-site DMRG algorithm gives a rough measure of the error associated with using a finite bond dimension, called the truncation error. Even though the truncation error is a crude approximation of the true error, protocols for extrapolating the truncation error to zero can give greatly improved energy estimates and approximate errors on those estimates. Another trick is to rely as much as possible on measurements of local quantities rather than long-distance correlation functions, which are determined by DMRG much less accurately. Similar information to correlation functions can be obtained from a local perturbation of the system followed by the measurement of local quantities away from the perturbation by following the linear response theory. Note that perturbations, e.g.\ a global antiferromagnetic field breaking the $\mathrm{SU}(2)$ symmetry of an antiferromagnet, can sometimes reduce entanglement, making the calculation easier.

The energy variance and the related V-score presented in this paper deviates from the usual DMRG protocols, but they provide a natural way to compare different algorithms. The calculation of the variance of the energy is straightforward in DMRG, but it is much more costly to compute than the truncation error. Aside from the cost, extrapolations to zero variance are a potential improvement over truncation error extrapolations. To mitigate the cost, a two-site variance has been introduced, which is potentially more robust than the truncation error, and can be used in the one-site DMRG algorithm~\cite{TwoSiteDMRG}. The two-site variance has a cost similar to the rest of a two-site DMRG sweep. Here, however, we only report full-variance calculations and forgo extrapolations to allow comparisons with other methods.

A key limitation in calculating the full variance is that one can run out of memory. We use a matrix product operator (MPO) form for the Hamiltonian, and calculating the variance can be done using the square of this MPO --- except that this tends to require large amounts of memory. A useful trick is to break the Hamilonian MPO into $k$ pieces, each with smaller bond dimension. Then one needs to sum $k^2$ terms which may be calculated in parallel.

\subsubsection{Fork tensor product states for impurity models}

The many-body wave function of the impurity models is parameterized by the Fork tensor product states (FTPS)~\cite{bauernfeind2017fork}. Compared to MPS, which has a chain geometry, FTPS avoids the artificial long-range interaction that is detrimental to MPS by explicitly separating bath degrees of freedom belonging to different bands. Hence, the FTPS is expected to efficiently capture the entanglement structure of multiorbital problems. Furthermore, the bipartite nature of the FTPS makes it possible to extend the efficient DMRG algorithm developed for MPS to FTPS. The ground state is found in our calculations by the single-site DMRG algorithm supplied with a subspace expansion~\cite{hubig2015strictly}. Except for the three-band model with spin-orbital coupling, which has only $\mathrm{U}(1)$ symmetry for the charge sector, all calculations are performed under the global $\mathrm{U}(1)$ symmetries for the charge and spin sectors. We first perform 30 DMRG sweeps without symmetries with a relatively low bond dimension to find the correct charge and spin sector. Then, the charge and the spin quantum numbers are fixed to be $\ev*{\hat{N}_\alpha}$ and $\ev*{\hat{S}^z_\alpha}$ for each spin-orbital $\alpha$, respectively. Finally, the ground state is found by another 60 DMRG sweeps in the fixed quantum number sector with a maximum bond dimension $m = 100$ for single-band models and $m = 350$ for three-band models.

\subsection{Variational Monte Carlo} \label{sec:VMC}

Variational Monte Carlo (VMC) methods are a family of computational methods that do not suffer from the sign problem and whose computational cost is tractable.
In particular, VMC combines a variational encoding of the wave function, to reduce the memory complexity associated with \emph{storing} the wave function, with Monte Carlo techniques which lower the computational complexity.
This approach was originally introduced to treat models in the continuum such as the helium atom or the electron gas~\cite{McMillan1965,Ceperley1977}, and later adapted to find the ground states of lattice systems as those discussed in the main text of this manuscript~\cite{Horsch1983,Shiba1986}.
Since then, more sophisticated methods have been proposed on various approximation levels~\cite{YokoyamaShiba1987,GrossJoyntRice1987,Capriotti2001,Tahara2008,MISAWA2019447,Nomura2017}.
Nowadays, several open-source software implementations of those algorithms are available, such as mVMC~\cite{MISAWA2019447} and NetKet~\cite{netket3:2021}.

In variational approaches such as VMC, a quantum state $\ket{\psi_\theta}$ is encoded into a parameterized function $\psi_\theta(x)$ usually referred to as the \emph{variational ansatz}.
This function takes vectors $x$ from a certain basis of the Hilbert space as input and outputs the complex wave function amplitudes such that
\begin{equation}
\ket{\psi_\theta} := \sum_x\psi_\theta(x) \ket{x}.
\label{eq:var-wf}
\end{equation}
If the function $\psi_\theta$ is fixed, then a quantum state is uniquely identified by the vector of variational parameters $\theta$.
Variational ansatzes are usually chosen such that the number of variational parameters $\theta$ increases at most only algebraically with the number of degrees of freedom in the system (in most cases linearly or quadratically).
This provides an exponential reduction in the memory complexity with respect to storing the full wave function, which is a vector in an exponentially large Hilbert space.

Even when the wave function is encoded into a small vector of a few variational parameters $\theta$, the expectation value of physical quantities involves two sums over the entire basis of the Hilbert space, leading to exponential computational complexity.
To work around this issue, expectation values are computed through statistical averages of stochastic estimators.
Instead of summing over the whole Hilbert space, only a few basis elements are selected through rigorous Monte Carlo sampling techniques such as Markov chain Monte Carlo (MCMC).
In practice, the expectation value of the energy over the state, $E_\theta = \ev{\hat{H}}{\psi_\theta}$, is approximated by the following expectation value~\cite{becca_sorella_2017}:
\begin{equation}
E_\theta \approx \mathbb{E}_{x \sim P{_\theta(x)}}\left[ E^\text{loc}_\theta(x) \right],
\end{equation}
where the samples $\{x\}$ are distributed according to the Born probability
\begin{equation}
P_\theta(x) := \frac{\abs{\psi_\theta(x)}^2}{\ip{\psi_\theta}},
\label{eq:def-born-prob}
\end{equation}
and
\begin{equation}
E^\text{loc}_\theta(x) := \sum_{x'} \mel{x}{\hat{H}}{x'} \frac{\psi_\theta(x')}{\psi_\theta(x)}
\label{eq:def-eloc}
\end{equation}
is called the \emph{local energy}. Assuming that the Hamiltonian $\hat{H}$ has only a polynomial number of non-zero entries $\mel{x}{\hat{H}}{x'}$ for every row $\bra{x}$, the computational complexity of this local estimator is algebraic.

In general, when using VMC to determine the ground state of a Hamiltonian $\hat{H}$ one updates the variational parameters $\theta$ according to the conjugate gradient of the energy~\cite{Amari1998,sorella1998green}.
If $\psi_\theta(x)$ has real-valued parameters or if it is holomorphic, the gradient is estimated using the formula
\begin{equation}
\frac{\partial E_\theta}{\partial \theta^*} \approx \mathbb{E}_{x \sim P_\theta(x)}\left[ \left( E_\theta^\text{loc}(x) - E_\theta \right) \frac{\partial \log \psi_\theta^*(x)}{\partial \theta^*} \right],
\end{equation}
where $\frac{\partial \log \psi_\theta(x)}{\partial \theta}$ is computed by automatic differentiation (AD)~\cite{baydin2018automatic} in modern softwares.
This estimator also has the useful property that $\frac{\partial E_\theta}{\partial \theta^*} = 0$ when the variational state hits an eigenstate of the Hamiltonian such as the ground state, meaning that the optimization will stop if convergence is reached.
For non-holomorphic ansatzes with complex-valued parameters, the formula gains a second term $\propto \frac{\partial \log \psi_\theta^*(x)}{\partial \theta}$.

The simplest first-order optimization scheme is the stochastic gradient descent (SGD):
\begin{equation}
\theta \gets \theta - \eta \frac{\partial E_\theta}{\partial \theta^*},
\end{equation}
where $\eta$ is a sufficiently small positive number called the learning rate.
More elaborate optimization schemes that involve accumulating momentums of the gradient, such as Adam~\cite{kingma2014adam}, can also be employed.

A more advanced optimization method that leverages information about the local geometry of the variational manifold~\cite{provost1980riemannian,stokes2020quantum} is often used.
This method, known as natural gradient~\cite{Amari1998} in the machine learning literature and as stochastic reconfiguration (SR)~\cite{sorella1998green} in the VMC literature, approximates the imaginary-time evolution by $\exp(-\tau \hat{H})$ for the Hamiltonian $\hat{H}$ with sufficiently large $\tau$ to reach the ground state.

After the stochastic optimization of the energy, the variational wave function can be improved further by applying the Lanczos operator $\hat{\bbone} + \alpha \hat{H}$ to $\ket{\psi_\theta}$ once or twice with the variational parameter $\alpha$ and is employed in some cases of the present benchmark.

\subsubsection{Physically motivated ansatzes}

In fermionic systems, which are one of the main focuses of this paper, the variational ansatz takes the form ${\ket{\psi} = \hat{\mathcal{C}} \ket{\phi}}$, where $\ket{\phi}$ is an uncorrelated fermionic state and $\hat{\mathcal{C}}$ denotes a generic correlator. Centering the discussion on systems of spinful electrons, the uncorrelated state $\ket{\phi}$ is given either by a Slater determinant~\cite{YokoyamaShiba1987,GrossJoyntRice1987,ferrari2022charge} or by a pair-product (PP) wave function (also known as geminal wave function)~\cite{Gross1988,Capriotti2001,Tahara2008,astrakhantsev2021broken}. The variational PP wave function is defined by
\begin{equation}
\phi(x) := \ip{x}{\phi} = \mel{\bigl.x\bigr.}{\left( \sum_{i j \sigma \sigma'} f_{i j}^{\sigma \sigma'} \hat{c}_{i \sigma}^\dagger \hat{c}_{j \sigma'}^\dagger \right)^{\frac{N_{\rm e}}{2}}}{\phi_0},
\label{eq:PPwf}
\end{equation}
where $\ket{\phi_0}$ is the vacuum state, $N_{\rm e}$ is the number of electrons and the pair amplitudes $f_{i j}^{\sigma \sigma'}$ form a matrix of $4 N^2$ variational parameters on $N$ sites, which depend on the spatial coordinates $i$ and $j$, and the spins $\sigma$ and $\sigma'$. The spin-dependence of $f_{i j}^{\sigma \sigma'}$ can be chosen such that either singlet or triplet components appear in the wave function. The PP state, in general, contains Slater determinants as a subset and offers higher flexibility and accuracy. It can accommodate the Hartree--Fock--Bogoliubov type wave function with magnetic, charge, and superconducting orders~\cite{Tahara2008}, and paramagnetic metals as well, in a unified and flexible fashion.

The correlator $\hat{\mathcal{C}}$ operates to save the exponentially large number of basis functions, and frequently used examples in the present benchmark are given by introducing and applying various physically adequate operators in $\hat{\mathcal{C}}$, such as
\begin{equation}
\hat{\mathcal{C}} := \hat{\mathcal{L}}^S \hat{\mathcal{L}}^q \hat{\mathcal{P}}^{\rm G} \hat{\mathcal{P}}^{\rm J_c} \hat{\mathcal{P}}^{\rm J_s}.
\label{eq:C}
\end{equation}
The correlation factors
\begin{align}
\hat{\mathcal{P}}^{\rm G} &:= \exp \left( \sum_i g_i^{\rm G} \hat{n}_{i \spinup} \hat{n}_{i \spindown} \right), \\
\hat{\mathcal{P}}^{\rm J_c} &:= \exp \left( \sum_{i < j} g_{i j}^{\rm J_c} \hat{n}_i \hat{n}_j \right), \\
\hat{\mathcal{P}}^{\rm J_s} &:= \exp \left( \sum_{i < j} g_{i j}^{\rm J_s} \hat{S}^z_i \hat{S}^z_j \right)
\end{align}
are the Gutzwiller factor~\cite{PhysRevLett.10.159}, the long-range Jastrow correlation factor~\cite{PhysRev.98.1479,capello2005}, and the long-range spin Jastrow correlation factor~\cite{PhysRevLett.60.2531}, with spatially dependent variational parameters $g_i^{\rm G}$, $g_{i j}^{\rm J_c}$, and $g_{i j}^{\rm J_s}$, respectively; $\hat{S}^z_i = \frac{1}{2} \left( \hat{n}_{i \spinup} - \hat{n}_{i \spindown} \right)$ is the $z$ component of the local spin operator.
In practice, the translational symmetry is often imposed on the Gutzwiller and the Jastrow factors in order to reduce the number of independent variational parameters.

Furthermore, since on finite sizes the exact ground state possesses all the symmetries of the Hamiltonian, while the variational state $\ket{\phi}$ may break them, the quantum number projections $\hat{\mathcal{L}}$ are also considered to restore the symmetries. In \Eq{C}, $\hat{\mathcal{L}}^S$ and $\hat{\mathcal{L}}^q$ are examples of projection operators which enforce the fixed total spin $S$ and momentum $q$, respectively~\cite{Tahara2008}, and $\hat{\mathcal{L}}^q$ is used in the case of periodic boundary conditions.

In the above example, the variational parameters are contained both in the correlator $\hat{\mathcal{C}}$ (i.e.\ $g_i^{\rm G}, g_{i j}^{\rm J_c}$, and $g_{i j}^{\rm J_s}$) and in the uncorrelated state $\ket{\phi}$ (i.e.\ $f_{i j}^{\sigma \sigma'}$). They are optimized to better approximate the ground state by lowering the numerically evaluated energy.
By fully optimizing $f_{i j}^{\sigma \sigma'}$ including the long-range part, it is known that not only the symmetry-broken Hartree--Fock--Bogoliubov states and simple single-particle noninteracting Slater determinants, but also correlated metallic states such as Tomonaga--Luttinger liquid can be represented~\cite{Anderson_1988,Kaneko_2013}.
In order to lower the computational cost by reducing the number of independent variational parameters, one can assume that $f_{i j}^{\sigma \sigma'}$ has a sublattice structure such that it depends only on the relative position vector ${\bm r}_i - {\bm r}_j$ and a sublattice index of site $j$ which is denoted as $\eta(j)$. Then one can rewrite it as $f_{\eta(j)}^{\sigma \sigma'}({\bm r}_i - {\bm r}_j)$. In many cases, one can employ variational states which do not break any translational symmetry and assume a fully translational invariance ($1 \times 1$ sublattice structure, in the case of 2D models), restricting to singlet pairings only (i.e.\ $f_{i j}^{\sigma \sigma'} = -f_{i j}^{\sigma'\!, \sigma} (1 - \delta_{\sigma \sigma'})$). Antiferromagnetic states can be included by extending the sublattice structure to a $2 \times 2$ (or larger) unit cell. In most of the present benchmark studies, we do not restrict the sublattice size; namely, the sublattice size is the same as the full lattice. In some cases, $\hat{\mathcal{P}}^{\rm J_s}$ and $\hat{\mathcal{L}}^S$ are omitted.
One can also improve the wave function by implementing dependence on the local density of $\ket{x}$ (see Ref.~\cite{Ido2022}).
The advantage of using this scheme together with Eqs.~\eqref{eq:PPwf} and \eqref{eq:C} is that the quantum entanglement beyond the area law can be taken into account.

In an alternative approach, the parameters $f_{i j}^{\sigma \sigma'}$ for $\ket{\phi}$ are obtained starting from an auxiliary non-interacting Hamiltonian, containing hopping and pairing amplitudes~\cite{Capriotti2001,Hu2013,Tocchio2019}. The most general form is given by
\begin{equation}
\hat{H}_\text{aux} := \sum_{i j \sigma \sigma'} \left( \chi_{i j}^{\sigma \sigma'} \hat{c}^\dagger_{i \sigma} \hat{c}_{j \sigma'} + \Delta_{i j}^{\sigma \sigma'} \hat{c}^\dagger_{i \sigma} \hat{c}^\dagger_{j \sigma'} + \text{h.c.} \right),
\label{eq:Haux}
\end{equation}
where $\chi_{i j}^{\sigma \sigma'}$ and $\Delta_{i j}^{\sigma \sigma'}$ are hopping and pairing terms, respectively. Then the ground state of $\hat{H}_\text{aux}$ is a PP wave function in \Eq{PPwf}, with the pair amplitudes $f_{i j}^{\sigma \sigma'}$ determined by the parameters $\chi_{i j}^{\sigma \sigma'}$ and $\Delta_{i j}^{\sigma \sigma'}$.
The advantage of this approach is two-fold. First, long-range pair amplitudes can be obtained within a short-range parametrization of $\chi_{i j}^{\sigma \sigma'}$ and $\Delta_{i j}^{\sigma \sigma'}$, thus avoiding delicate optimizations of the long-range tails of $f_{i j}^{\sigma \sigma'}$.
Second, symmetries may be imposed directly on the auxiliary Hamiltonian, thus avoiding including further projectors, e.g., $\hat{\mathcal{L}}^S$ and $\hat{\mathcal{L}}^q$. For example, by restricting to ``singlet'' hoppings ($\chi_{i j}^{\sigma \sigma'} = \delta_{\sigma \sigma'} \chi_{i j}$ with $\chi_{i j} = \chi_{j i}^*$) and pairings ($\Delta_{i j}^{\sigma \sigma'} = (1 - \delta_{\sigma \sigma'}) \Delta_{i j}$ with $\Delta_{i j} = \Delta_{j i}$), the PP wave function is already a singlet, while by imposing translational symmetry in the dependence of these parameters on the lattice sites, the PP state is translationally symmetric automatically. Within this framework, the magnetic order can be described by including on-site hoppings which mimic the presence of a (site-dependent) Zeeman field $\bm{h}_i = (h^x_i, h^y_i, h^z_i)$, with $h^z_i = \frac{1}{2} (\chi_{i i}^{\spinup \spinup} - \chi_{i i}^{\spindown \spindown})$, $h^x_i = \chi_{i i}^{\spinup \spindown} + \chi_{i i}^{\spindown \spinup}$, and $h^y_i = \mathrm{i} (\chi_{i i}^{\spinup \spindown} - \chi_{i i}^{\spindown \spinup})$ (see for example Refs.~\cite{Iqbal2016,Ferrari2018}).

The general parametrization in \Eq{Haux} allows us to include site-dependent terms to describe charge and/or spin inhomogeneities (i.e.\ stripes)~\cite{Tocchio2019,Marino2022}. Jastrow factors, when included, are chosen to have translational and rotational symmetries. Furthermore, backflow correlations can be included, redefining the orbitals of \Eq{Haux} on the basis of the many-body electronic configuration~\cite{Tocchio2008,Tocchio2011}. Analogous variational wave functions can be employed to study localized spins systems, e.g., frustrated Heisenberg models on different lattice geometries~\cite{Capriotti2001,Hu2013,Iqbal2013,Iqbal2016,Ferrari2019}.

Notice that in the context of spin models, the Gutzwiller factor is replaced by the Gutzwiller projector $\hat{\mathcal{P}}^{\rm G}_\infty := \prod_i \hat{n}_i (2 - \hat{n}_i)$, which enforces single occupancy on each lattice site, on top of a fermionic uncorrelated PP state $\ket{\phi}$. In addition, analogously to the electronic case previously discussed, a spin Jastrow factor $\hat{\mathcal{P}}^{\rm J_s}$ can be included, and quantum number projectors $\hat{\mathcal{L}}^S$, $\hat{\mathcal{L}}^q$ can be applied to enforce lattice symmetries. Also, in this case, the $f_{i j}^{\sigma \sigma'}$ pair amplitudes of the PP state can be assumed to be direct variational parameters (with certain symmetry constraints) or defined through an auxiliary Hamiltonian like the one of \Eq{Haux}.

In addition, we mention that a possible bias inevitable in the original wave function can be progressively removed by adding a correlator $\hat{\mathcal{M}}$ implemented as a restricted Boltzmann machine (RBM)~\cite{carleo_solving_2017,Nomura2017} on top of the variational state, i.e., $\ket{\psi} = \hat{\mathcal{C}} \hat{\mathcal{M}} \ket{\phi}$. In this case, Jastrow factors can be omitted to save computational cost. We refer the reader to the following subsection for more details about the RBM. Recently, the VMC approach with the RBM implementation has been successfully applied to reveal the nature of the quantum spin liquids~\cite{ferrari2019neural,Nomura2021,Ido2022}.

\subsubsection{Neural quantum states}

Neural networks are structured nonlinear functions that can universally approximate any well-behaved function. Motivated by their recent success of efficiently representing probability distributions in machine learning tasks, they have also been used as variational ansatzes to represent quantum states. They typically have more parameters and higher expressive power than traditional ansatzes.
The first proposed neural quantum state (NQS) is a restricted Boltzmann machine (RBM)~\cite{carleo_solving_2017} applied first to quantum spin Hamiltonians and then extended to fermionic systems~\cite{Nomura2017} as is introduced in the last subsection. The RBM is also equivalent to a multilayer perceptron (MLP) with two layers.
The finite temperature path integral formalism was shown to exactly map to a deep Boltzmann machine (DBM), where the Trotter--Suzuki layers correspond to the multiple hidden layers in the DBM~\cite{CarleoDBM_2018}.
The correlator $\hat{\mathcal{M}}$ represented by NQS can also benefit from techniques developed for traditional ansatzes, such as backflow correlation~\cite{luo2019backflow} and Gutzwiller projection~\cite{ferrari2019neural}.

Inspired by the development in computer vision, NQS based on convolutional neural networks (CNN)~\cite{choo2019two} takes advantage of the locality and the translational symmetry of physical systems on regular lattices.
We can also restore symmetries by quantum number projections discussed in the previous subsection.
Some of the data in this paper are obtained by RBM combined with quantum number projections~\cite{Nomura_Helping_2021}.
Group convolutions~\cite{luo2021gauge,roth2021group} further generalize them to richer symmetries including rotations, reflections, and gauge transformations.

Another direction for developing NQS is perfect sampling from complicated many-body probability distributions, which avoids the issue of autocorrelation time in MCMC. It can be achieved by autoregressive neural networks (ARNN)~\cite{sharir2020deep}, which are exact likelihood models. They rely on decomposing a joint probability into conditional probabilities that can be sequentially sampled, and they can comprise either dense or convolutional layers. Recurrent neural networks (RNN)~\cite{hibat2020recurrent,Roth2020} maintain a memory of past information during the sampling procedure, which makes them suitable to capture long-range correlations, as suggested by their success in natural language processing.

One interesting feature about the RNN is its capability to encode translation-invariant properties of the bulk of a quantum system~\cite{Roth2020}. Moreover, RNNs can be extended to multiple spatial dimensions. In particular, one can construct 2D RNNs that were shown to be competitive with DMRG, and also cheaper in terms of computational complexity compared to projected-entangled pair states (PEPS)~\cite{hibat2020recurrent}. RNNs can use much fewer parameters than other architectures to encode information in a large spatial area, thanks to weight sharing between RNN cells at all sites.
In a similar fashion to tensor networks, tensorized versions of RNNs have been used to provide higher expressive power~\cite{VNA2021,TensorizedRNNs2021,Wu2022} and more accurate estimations of ground-state energies that outperform DMRG in certain regimes~\cite{TensorizedRNNs2021,Wu2022}. They were also shown to be able to encode the area law of entanglement~\cite{Wu2022}. Symmetries, such as $\mathrm{U}(1)$ symmetry and spatial symmetries, can be imposed in RNNs to improve the quality of variational calculations~\cite{hibat2020recurrent,TensorizedRNNs2021}. Being the recent state of the art for many machine learning tasks, transformers have also been proposed for NQS as they have a more flexible autoregressive architecture~\cite{luo2021gauge,zhang2022transformer}.
Additionally, it is worth noting that thermal-like fluctuations can be added to the training of autoregressive models in the hope of escaping local minima that can be encountered when studying disordered or frustrated systems~\cite{Roth2020,VNA2021,TensorizedRNNs2021}.

Neural networks can be used as well to simulate fermionic systems within the second quantization formalism~\cite{Choo2020chemistry,Yoshioka2021solids,Bennewitz2022neuralErrorMitigation}. The fermionic modes are mapped into an interacting quantum spin model in this formalism. This can be achieved via the Jordan--Wigner~\cite{Jarodan1928JW}, the parity, or the Bravyi--Kitaev~\cite{Bravyi2002} transformations. The reduction of the fermionic problem into a spin Hamiltonian makes it possible to exploit the success of NQS on spin systems. However, this approach suffers from the disadvantage of producing a spin Hamiltonian with non-local interactions. First quantization is therefore an attractive alternative as it preserves the locality of the physical interactions. In this case, the amplitudes of the variational state $\psi_\theta(x)$ must be antisymmetric with respect to permutations of the particle indices. NQS-based parametrizations of fermionic wave functions borrow inspiration from traditional ansatzes like the Slater--Jastrow state, backflow correlations, and hidden particle representations.

The amplitudes of a Slater--Jastrow-inspired NQS consist of the product of a parameterized antisymmetric reference factor $\ket{\phi}$ (Slater determinant or PP) and a symmetric neural network factor $\hat{\mathcal{M}}$. The neural network is in charge of incorporating correlations on top of the reference wave function. First introduced in Ref.~\cite{Nomura2017}, a positive RBM was used as the correlation factor. Later works also implemented correlation factors that can alter the nodal structure of the reference state and respect the translational symmetry via the use of CNN with skip connections~\cite{Stokes2020spinless}. It it noted that if $\ket{\phi}$ is a general Slater determinant of non-orthogonal orbitals, then the NQS is a universal parametrization in the lattice~\cite{RobledoMoreno2022hidden}. Alternatively, neural networks have been used to parametrize the $N$-particle orbitals of an $N$-particle determinant in order to encode backflow correlations. This variational family has also been shown to be universal in the lattice~\cite{luo2019backflow}.

Lastly, antisymmetric NQS has also been constructed using ``hidden'' additional fermionic degrees of freedom~\cite{RobledoMoreno2022hidden}. In this case, the variational state is represented by a Slater determinant in the Hilbert space that is augmented by adding the ``hidden'' fermions. The Slater determinant in the augmented space is then projected into the physical Hilbert space, and this projection is parameterized by a neural network. The neural network parameters are optimized together with the single-particle orbitals of the determinant. This ansatz explicitly contains the above Slater--Jastrow-inspired factorization, as well as a compact representation of configuration-interaction wave functions~\cite{RobledoMoreno2022hidden}.

\subsubsection{Variational auxiliary-field quantum Monte Carlo}

The variational auxiliary-field quantum Monte Carlo (VAFQMC)~\cite{Sorella_Arxiv2022} approach creates a variational ansatz for the ground state wave function of the Hubbard Hamiltonian $\hat{H} = \hat{K} + \hat{V}$, by projections via an optimizable pseudo-Hamiltonian, using the formalism of AFQMC (see Section~\ref{sec:AFQMC}). In VAFQMC, a single Slater determinant $\ket{\psi_\text{MF}}$ is first constructed from an effective mean-field calculation with a set of variational parameters $\bm{\alpha}_0$, such that $\hat{H}_\text{MF}(\bm{\alpha}_0) \ket{\psi_\text{MF}} = E_0(\bm{\alpha}_0) \ket{\psi_\text{MF}}$. A variational ansatz is then constructed, by operating onto $\ket{\psi_\text{MF}}$ a projection operator:
\begin{equation}
\ket{\psi_{\tau}} = \exp\qty(-\frac{\tau}{2} \qty(\hat{H}_\text{MF}(\bm{\alpha}) + \hat{V})) \ket{\psi_\text{MF}},
\label{eq:ansatz}
\end{equation}
where $\bm{\alpha}$ denotes a set of variational parameters.

In \Eq{ansatz}, $\hat{H}_\text{MF}(\bm{\alpha})$ replaces the kinetic part of the Hubbard Hamiltonian with a general quadratic operator of fermionic creation and annihilation operators, which can include, for instance, a $d$-wave BCS pairing field. In this work, $\hat{H}_\text{MF}(\bm{\alpha})$ was designed to give a $\ket{\psi_\text{MF}}$ describing AFM stripes with the expected wavelength~\cite{Xu_PRR2022}. The potential part $\hat{V}$ is kept as the original Hubbard on-site interaction. The parameter $\tau$, the total imaginary time of the projection, is kept fixed and plays the role of an effective inverse temperature.

The projection in \Eq{ansatz} is further broken up as
\begin{align}
& \exp\left( -\frac{\tau}{2} \left( \hat{H}_\text{MF}(\bm{\alpha}) + \hat{V} \right) \right) = \nonumber \\
& \quad \prod_{i = 1}^n \exp\left( -t_i \hat{H}_\text{MF}(\bm{\alpha}) \right) \exp\left( -h_i \hat{V} \right) \nonumber \\
& \quad \times \exp\left( -t_{n + 1} \hat{H}_\text{MF}(\bm{\alpha}) \right),
\end{align}
where $n$ is the number of time steps given by $n = \max\left( [(U \tau / 0.4 - 1) / 5], 1 \right)$ depending on the interaction strength $U$. The variable steps $h_i$ and $t_i$ are treated as additional variational parameters to optimize in order to minimize Trotter errors~\cite{Beach_PRB2019}. However, here we introduce a simple functional form for the non-uniform time steps which depends on a single variational parameter $\Delta \tau$ (see Ref.~\cite{Sorella_Arxiv2022} for details). The variational ansatz $\ket{\psi_{\tau}}$ in \Eq{ansatz} can thus be equivalently denoted as $\ket{\psi_n}$, yielding a variational energy $E_n := \ev{\hat{H}}{\psi_n} / \ip{\psi_n}$.

Now following similar procedures to AFQMC, we recast the variational energy as
\begin{equation}
E_n = \frac{\sum_{\bm{\sigma} \bm{\sigma}'} \ev{\hat{U}_n^\dagger(\bm{\sigma}') \hat{H} \hat{U}_n(\bm{\sigma})}{\psi_\text{MF}}}{\sum_{\bm{\sigma} \bm{\sigma}'} \ev{\hat{U}_n^\dagger(\bm{\sigma}') \hat{U}_n(\bm{\sigma})}{\psi_\text{MF}}},
\label{eq:expectation}
\end{equation}
where
\begin{align}
\hat{U}_n(\bm{\sigma}) &:= \prod_{i = 1}^n \exp\left( -t_i \hat{H}_\text{MF}(\bm{\alpha}) \right) \exp\Bigl( \lambda_i \sum_j \sigma_{j, i} \hat{m}_j \Bigr) \nonumber \\
&\phantom{:={}}\times \exp\left( -t_{n + 1} \hat{H}_\text{MF}(\bm{\alpha}) \right),
\end{align}
with the vector $\bm{\sigma} := \{\sigma_{j i}\} \in \{\pm 1\}^{N_\text{s} \times n}$ denoting the collection of auxiliary fields arising from the discrete Hubbard--Stratonovich transformation~\cite{Hirsch_PRB1985}. The spin operator for each site $j$ is defined as $\hat{m}_j = \hat{n}_{j, \spinup} - \hat{n}_{j, \spindown}$, and the parameter $\lambda_i$ is given by $\cosh \lambda_i = \exp(U h_i)$.

\Eq{expectation} is now rewritten in a form suitable for Monte Carlo:
\begin{equation}
E_n = \frac{\sum_{\bm{\sigma} \bm{\sigma}'} \abs{W_n(\bm{\sigma}', \bm{\sigma})} e_n(\bm{\sigma}', \bm{\sigma}) S_n(\bm{\sigma}', \bm{\sigma})}
{\sum_{\bm{\sigma} \bm{\sigma}'} \abs{W_n(\bm{\sigma}', \bm{\sigma})} S_n(\bm{\sigma}',\bm{\sigma})},
\end{equation}
where
\begin{align}
W_n(\bm{\sigma}', \bm{\sigma}) &:= \ev{\hat{U}_n^\dagger(\bm{\sigma}') \hat{U}_n(\bm{\sigma})}{\psi_\text{MF}}, \\
S_n(\bm{\sigma}', \bm{\sigma}) &:= \frac{W_n(\bm{\sigma}', \bm{\sigma})}{\abs{W_n(\bm{\sigma}', \bm{\sigma})}}, \\
e_n(\bm{\sigma}', \bm{\sigma}) &:= \frac{\ev{\hat{U}_n^\dagger(\bm{\sigma}') \hat{H} \hat{U}_n(\bm{\sigma})}{\psi_\text{MF}}}{W_n(\bm{\sigma}', \bm{\sigma})}
\end{align}
are the weight factor, the phase factor, and the local energy respectively. For each set of variational parameters, we estimate the energy by $E_n = \langle S_n e_n \rangle_{W_n} / \langle S_n \rangle_{W_n}$, where $\langle\,\cdot\,\rangle_{W_n}$ indicates the average of a random variable with respect to the Monte Carlo samples from the probability distribution $|W_n|$.

The optimization of the variational parameters is then carried out by generalizing techniques from standard variational Monte Carlo~\cite{becca_sorella_2017,sorella1998green} and from machine learning~\cite{Amari1998}. Minimizing $E_n(\bm{\alpha})$ with respect to $\bm{\alpha}$ requires the energy derivatives. Assuming there are $2 p$ variational parameters $\alpha_1, \alpha_2, \ldots, \alpha_{2 p}$ plus an extra parameter of the minimum time step $\alpha_{2 p + 1} = \Delta \tau$, we can compute the derivatives by
\begin{equation}
\frac{\partial E_n}{\partial \alpha_j} = \frac{\left\langle S_n \qty(\frac{\partial e_n}{\partial\alpha_j} + (e_n - E_n) O_j) \right\rangle_{W_n}}{\left\langle S_n \right\rangle_{W_n}},
\end{equation}
where $O_j := \frac{\partial \ln W_n}{\partial \alpha_j}$. The complex derivatives $\frac{\partial e_n}{\partial\alpha_j}$ and $O_j$ are obtained by automatic differentiation~\cite{Griewank_SIAM2008}. The final optimized $E_n$ defined within the variational ansatz is dependent on the parameter $\tau$, and provides an upper bound to the true ground-state energy of the given Hamiltonian.

\subsection{Parameterized quantum circuits and variational quantum eigensolver}

Variational states can be represented by parameterized quantum circuits (PQC)~\cite{cerezo2021variational}. A PQC has parameterized single and two-qubit gates, e.g.\ rotation gates, as well as multi-qubit entangling gates~\cite{nielsen_chuang_2010}.
Different hardware architectures come with different sets of native gates.
The set of variational parameters $\theta$ is optimized using the variational quantum eigensolver (VQE) algorithm~\cite{Peruzzo_2014}. When using quantum hardware, the expectation value of the Hamiltonian and the energy gradient components must be measured through repeated wave function collapses, thus introducing a sampling noise similar to the one in VMC simulations. We only consider noiseless emulations of the quantum circuits, i.e., neglecting this measurement sampling noise, as well as hardware noise due to decoherence. We do this because our goal is to test whether the relation between the V-score and the relative energy accuracy also holds in the quantum computing setting.

We employ several types of PQC ansatzes and optimization methods.
We first tackle the TFIM at the criticality $\Gamma = 1$, defined on a finite chain with $L$ sites.
The circuit is made of a series of $d$ blocks built from single-qubit rotations $\hat{U}_\text{R}(\theta^k)$, interlayered with entangler blocks $\hat{U}_\text{ent}$, that spans the required length of the qubit register, with $k = 1, \ldots, d + 1$.
This is made of a ladder of CNOT (also known as CX) gates with linear connectivity, such that qubit $q_i$ is target of qubit $q_{i - 1}$ and controls qubit $q_{i + 1}$ for $i = 1, \ldots, L - 2$.
Since the single-qubit rotations are all local operations, $\hat{U}_\text{R}(\theta^k)$ can be written as a tensor product of rotations on single qubits:
\begin{equation}
\hat{U}_\text{R}(\theta^k) = \bigotimes_{i = 0}^{L - 1} \hat{R}_y(\theta^k_{q_i}),
\end{equation}
where $\hat{R}_y(\theta^k_{q_i})$ is a rotation around the y-axis on the Bloch sphere of qubit $q_i$.
The full unitary circuit operation is described by
\begin{equation}
\hat{U}_\text{R-CX}(\theta) := \hat{U}_\text{R}(\theta^{d + 1}) \prod_{i = d}^1 \hat{U}_\text{ent} \hat{U}_\text{R}(\theta^i),
\label{eq:cnot-ansatz}
\end{equation}
and the final parameterized state is
\begin{equation}
\ket{\psi(\theta)} = \hat{U}_\text{R-CX}(\theta) \ket{0}^{\otimes L}.
\label{eq:trial-ansatz-cnot}
\end{equation}
The accuracy of the calculation is controlled by the circuit depth $d$, and the total number of variational parameters is $L (d + 1)$.

An alternative is to use physically motivated ansatzes, such as the Hamiltonian variational (HV) ansatz~\cite{wecker2015progress}.
The unitary operator defining the HV ansatz is made of $d$ blocks, and each block is a product of $\ell$ operators $\hat{U}_\alpha = \exp(\mathrm{i} \theta_\alpha^k \hat{H}_\alpha)$, with $\alpha = 1, \ldots, \ell$ indexing the non-commuting terms of the Hamiltonian. For TFIM we only need $\ell = 2$. In this case, the full unitary operator is
\begin{equation}
\hat{U}_\text{HV}(\theta) := \prod_{i = d}^1 \hat{U}_2(\theta^i_2) \hat{U}_1(\theta^i_1),
\label{eq:hv-ansatz}
\end{equation}
which can be efficiently decomposed using one- and two-qubit quantum gates,
and the final parameterized state is
\begin{equation}
\ket{\psi(\theta)} = \hat{U}_\text{HV}(\theta) \left( \frac{\ket{0} + \ket{1}}{\sqrt{2}} \right)^{\otimes L},
\label{eq:trial-ansatz-hv}
\end{equation}
where the initial non-entangled state can be obtained from $\ket{0}^{\otimes L}$ by placing one Hadamard gate on each qubit.
The total number of parameters is $\ell d$.

\begin{figure}[htb]
\includegraphics[width=\linewidth]{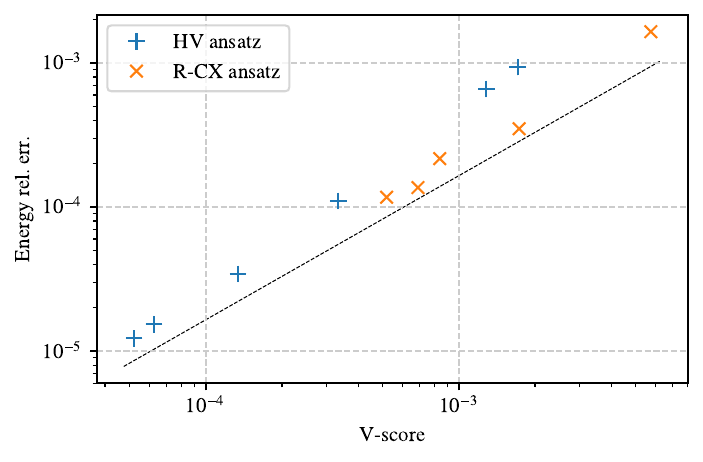}
\caption{\textbf{Comparison of V-scores of VQE ansatzes versus energy relative errors on a 10 sites TFIM.}
The black dashed line is the least square fit of all variational methods shown in Fig.~2 in the main text.
Each data point corresponds to a full VQE optimization using different circuit depths $d$.
For the HV ansatz, we report $d = 8, 12, 16, 20, 24, 26$, while for R-CX we report $d = 4, 6, 8, 10, 12$.
Longer circuits provide systematically better V-scores and energy relative errors.
}
\label{fig:v-score-rel-err-tfim-vqe}
\end{figure}

We use both the heuristic R-CX ansatz and the physically motivated HV ansatz on the TFIM. We have obtained a family of optimized trial states that depend on the circuit depth $d$, as shown in Fig.~\ref{fig:v-score-rel-err-tfim-vqe}, confirming the linear scaling of V-score versus energy relative error proposed in the main text.

For the Heisenberg and the {\JoneJtwo} models, we employ the symmetry-enhanced architecture introduced in Ref.~\cite{Seki_2020}. This PQC is more sophisticated compared to the previous ones, which are included mainly for demonstration purposes, and allows one to enforce translational, point-group, and $\mathrm{SU}(2)$ symmetries of the variational wave function in the device noise-free case. This circuit has shown outstanding performance in representing the ground state with only a few variational parameters when applied to 2D frustrated magnets~\cite{https://doi.org/10.48550/arxiv.2205.06278}.

The initial state to be prepared is the \emph{dimerized state}, splitting the $L$ site indices into $L / 2$ arbitrary pairs:
\begin{equation}
\ket{\psi_D} := \bigotimes_{0 \le i < L / 2} \frac{1}{\sqrt{2}} \Bigl( \ket{\spinup_{2 i} \spindown_{2 i + 1}} - \ket{\spindown_{2 i} \spinup_{2i + 1}} \Bigr),
\label{eq:dimerized}
\end{equation}
which is a product of $L / 2$ dimers on the selected pairs. This state is manifestly the $\mathrm{SU}(2)$-singlet. To access other total spin quantum numbers, one should replace one or more singlets in the product \Eq{dimerized} with triplets $\frac{1}{\sqrt{2}} \Bigl( \ket{\spinup_{2 i} \spindown_{2 i + 1}} + \ket{\spindown_{2 i} \spinup_{2 i + 1}} \Bigr)$.

We then introduce the \emph{SWAP operator}
\begin{equation}
\hat{P}_{i j} := \frac{1}{2} \left( \hat{\bm{S}}_i \cdot \hat{\bm{S}}_j + \hat{\bbone} \right),
\end{equation}
which exchanges spin states between sites $i$ and $j$. We note that the SWAP operator commutes with the total spin operator $\hat{S}^2$, therefore the variational ansatz
\begin{equation}
\ket{\psi(\theta)} = \left( \prod_\alpha \mathrm{e}^{\mathrm{i} \theta_\alpha \hat{P}_{i_\alpha j_\alpha}} \right) \ket{\psi_D}
\label{eq:wave-function}
\end{equation}
has a well-defined total spin quantum number. The exponential SWAP (eSWAP) operator $\mathrm{e}^{\mathrm{i} \theta_\alpha \hat{P}_{i_\alpha j_\alpha}}$ can be efficiently implemented on a quantum computer, as the two-qubit SWAP gate can be compiled into single-qubit gates and CNOTs, or is a native gate in architectures alternative to superconductive qubits.

Let us now suppose that, in addition to the $\mathrm{SU}(2)$ symmetry, the system also has translational or point-group symmetries, which can be all represented as qubit permutations. In Ref.~\cite{Seki_2020}, the authors provide a way that allows one to effectively project the wave function onto any irreducible representation subspace of the spatial symmetry. To this end, let us consider the spatial symmetry projector
\begin{gather}
\hat{P} := \frac{1}{|G|} \sum_{g \in G} \chi_g \hat{g},
\label{eq:symmetrization}
\end{gather}
where $G$ is the full spatial symmetry group consisting of the elementary unitary permutations $\hat{g}$, and $\chi_g$ are the characters depending on the desired projection quantum number (irreducible representation). The projected wave function $\ket{\psi_P(\theta)} = \frac{\hat{P}}{\sqrt{\mathcal{N}(\theta)}} \ket{\psi(\theta)}$ is normalized with $\mathcal{N}(\theta) := \ev{\hat{P}}{\psi(\theta)}$, since $\hat{P}^2 = \hat{P}$.

This wave function is optimized using the \emph{natural gradient} approach discussed in Section~\ref{sec:VMC}. The energy gradient is preconditioned using the \emph{metric tensor}, which mimics imaginary-time evolution in the allowed variational subspace. For this symmetry-enhanced ansatz, the energy gradient reads
\begin{equation}
\partial_i \ev*{E(\theta)} = 2 \Re\!\left( \frac{\mel{\psi(\theta)}{\hat{H} \hat{P}}{\partial_i \psi(\theta)}}{\mathcal{N}(\theta)} - \mathcal{A}_i(\theta) \ev*{E(\theta)} \right),
\label{eq:gradient}
\end{equation}
where $\mathcal{A}_i(\theta) := \frac{1}{\mathcal{N}(\theta)} \mel{\psi(\theta)}{\hat{P}}{\partial_i \psi(\theta)}$ is the \emph{connection}. The metric tensor is defined as
\begin{equation}
G(\theta)_{i j} := \frac{\mel{\partial_i \psi(\theta)}{\hat{P}}{\partial_j \psi(\theta)}}{\mathcal{N}(\theta)} - \mathcal{A}^*_i(\theta) \mathcal{A}_j(\theta).
\label{eq:metric-tensor}
\end{equation}
The parameter update reads
\begin{equation}
\theta \gets \theta - \eta \sum_j \bigl( \Re G(\theta) \bigr)^{-1}_{i j} \partial_j \ev*{E(\theta)}.
\end{equation}

The matrix elements required to construct these objects can be measured using the Hadamard test rule~\cite{Seki_2020,https://doi.org/10.48550/arxiv.2205.06278}. During the simulation of the VQE optimization on a classical computer, we either measure these matrix elements using $N_s$ circuit shots or compute them exactly. In the former case, the metric tensor obtained with sampling should be regularized to make the matrix inversion well defined. Here, we employ the regularization $G_\text{reg} = \sqrt{G G} + \beta \bbone$ as suggested in Ref.~\cite{gacon2021simultaneous}.

Finally, it is also important to check that the V-score can be efficiently computed with VQE.
The current way to estimate the energy with VQE is by calculating the weighted sum of the expectation values of all Pauli operators that compose the Hamiltonian.
For instance, the TFIM is made of $L$ local terms, yet most of them can be measured simultaneously~\cite{kandala_hardware-efficient_2017}. It turns out that only two bases are needed: the computational basis, and the rotated $\{\hat{\sigma}^x\}^{\otimes L}$ basis.
A similar argument applies for $\hat{H}^2$: While an upper bound for the number of terms is $L^2$, it is possible to check that the number of groups of Pauli operators that can be measured simultaneously grows sub-linearly with the system size, in both one and two dimensions.

The 2D Hubbard model introduces the additional complication of the fermion-to-qubit mapping. In more than one dimension, the Jordan--Wigner mapping generally transforms a two-local fermionic operator, i.e.\ the hopping term, into a non-local qubit operator~\cite{Peruzzo_2014}.
However, numerical tests show that the number of bases, thus the number of measurements, grows only sub-quadratically with the system size.

\subsection{Quantum Monte Carlo}

In the benchmarks, we include some quantum Monte Carlo (QMC) methods. Although they are not strictly variational and the V-score is not applicable to them, they produce numerically exact results in some cases, while in other cases highly accurate energies and the bias has been certified to be small. These cases are discussed below, with the methods described in detail. The QMC results are used to assess the V-scores of variational methods when ED is not practical.

\subsubsection{Auxiliary-field quantum Monte Carlo} \label{sec:AFQMC}

There exist two different auxiliary-field Monte Carlo algorithms, based on one hand the finite temperature grand canonical ensemble~\cite{Blankenbecler_1981}, and on the other hand the ground state canonical ensemble~\cite{Sorella_1989,Imada_1989}. The auxiliary-field quantum Monte Carlo (AFQMC) method (for an overview, see e.g.\ Ref.~\cite{Shiwei_LectureNotes2019}) used in this work is based on the latter.
It filters out the ground state from an initial state by an imaginary-time propagation: $\ket{\psi_G} = \lim_{\tau \rightarrow \infty} \mathrm{e}^{-\tau \hat{H}} \ket{\psi_I}$, where $\ket{\psi_I}$ and $\ket{\psi_G}$ are the initial and ground state respectively, $\tau$ the imaginary time, and $\hat{H}$ the many-body Hamiltonian. The initial state must satisfy $\braket{\psi_I}{\psi_G} \neq 0$ but can be otherwise arbitrary. We have typically taken it from a mean-field calculation~\cite{Mingpu_PRB2016,Purwanto_JCP2008}.

The projection is achieved by first discretizing the imaginary time into $m$ small time steps $\Delta \tau$: $\mathrm{e}^{-\tau \hat{H}} = (\mathrm{e}^{-\Delta \tau \hat{H}})^m$,
and then applying Trotter-Suzuki breakup~\cite{Trotter_1959,Suzuki_1976} to each time step:
\begin{equation}
\mathrm{e}^{-\Delta \tau \hat{H}} = \mathrm{e}^{-\Delta \tau \hat{K} / 2} \mathrm{e}^{-\Delta \tau \hat{V}} \mathrm{e}^{-\Delta \tau \hat{K} / 2} + O\qty(\Delta \tau^3).
\label{eq:Trotter}
\end{equation}
Here $\hat{K}$ is the kinetic part consisting of one-body operators, and $\hat{V}$ the interacting part containing two-body operators. For all systems computed in this work, we have either extrapolated $\Delta \tau$ to zero or set it to a fixed value (typically $\Delta \tau = 0.01$) and verified that the Trotter error is well within our statistical error.

Hubbard--Stratonovich (HS) transformation is applied to rewrite the interacting part into a one-body form coupled with auxiliary fields (AF) $\{x_i\}$. In the Hubbard model, a spin decomposition is applied to the $U > 0$ case:
\begin{equation}
\mathrm{e}^{-\Delta \tau U \hat{n}_{i \spinup} \hat{n}_{i \spindown}} = \mathrm{e}^{-\Delta \tau U (\hat{n}_{i \spinup} + \hat{n}_{i \spindown}) / 2} \sum_{x_i = \pm 1} \frac{1}{2} \mathrm{e}^{\gamma x_i (\hat{n}_{i \spinup} - \hat{n}_{i \spindown})},
\label{eq:spin-HS}
\end{equation}
where $\cosh \gamma = \exp(\Delta \tau U / 2)$. For the $U < 0$ cases, we apply a charge decomposition form of the HS transformation. A systematic study of the effect of the different HS transformations can be found in Ref.~\cite{Hao_PRB2013}.

The projection is carried out by evaluating the propagator as an integral using the Monte Carlo method:
\begin{equation}
\mathrm{e}^{-\Delta \tau \hat{H}} = \int p(\bm{x}) \hat{B}(\bm{x}) \dd \bm{x},
\end{equation}
where $\bm{x} := \{x_i\} = \{x_1, x_2, \ldots, x_{N_\text{s}}\}$ for a lattice with $N_\text{s}$ sites, and $p(\bm{x})$ is a probability distribution in AF space, which is a uniform function in the discrete HS transformation above.
The propagator $\hat{B}(\bm{x})$ now only contains one-body operators: $\hat{B}(\bm{x}) := \mathrm{e}^{-\Delta \tau \hat{K} / 2} \hat{\bm{b}}(\bm{x}) \mathrm{e}^{-\Delta \tau \hat{K} / 2}$, where $\hat{\bm{b}}(\bm{x}) := \prod_{i = 1}^{N_\text{s}} \hat{b}_i(x_i)$ is the product of the one-body operators transformed from the interacting part, i.e., the right-hand side of \Eq{spin-HS}.

In this work, two different ground state AFQMC methods are used. For sign-problem-free systems, e.g., the repulsive Hubbard model ($U > 0$) at half-filling or the spin-balanced attractive Hubbard model ($U < 0$)~\cite{Loh_PRB1990,Hirsch_PRB1985}, numerically exact results are given by AFQMC using a generalized Metropolis algorithm with force bias~\cite{Hao_PRA2015,Hao_PRE2016}. On the other hand, for a doped Hubbard model with $U > 0$, AFQMC is applied with a constraint path (CP) approximation~\cite{Shiwei_PRB1997,NGUYEN_CPC2014} to control the sign problem. The two ground state AFQMC methods are described in the following two subsections, respectively.

\subsubsection{Sign-problem-free Hubbard model: exact AFQMC with generalized Metropolis algorithm}

For the half-filled repulsive Hubbard model on a square lattice, as well as the spin-balanced attractive Hubbard model, the sign problem is absent in AFQMC because of symmetry. In these cases, exact calculations are performed.
(Note that in almost all sign-problem-free calculations, the standard approach has an infinite variance problem which must be properly taken care of~\cite{Hao_PRE2016}.)
We study these systems using the AFQMC with a generalized Metropolis algorithm~\cite{Hao_PRA2015}.
The ground-state energy is given by $\ev*{\hat{H}} = \mel{\psi_L}{\hat{H}}{\psi_R} / \ip{\psi_L}{\psi_R}$, where $\bra{\psi_L} := \bra{\psi_I} \exp\mqty(-\tau_L \hat{H})$ and $\ket{\psi_R} := \exp\mqty(-\tau_R \hat{H}) \ket{\psi_I}$.
Since the Hamiltonian commutes with propagators, energy measurements can be inserted between any two time steps, i.e., with any combination of $\tau_L + \tau_R = \tau$, provided $\tau$ is sufficiently larger than the equilibration time, $\tau > \tau_\text{eq}$, to reach the ground state from $\ket{\psi_I}$.

To illustrate the sampling process, we rewrite the energy expectation as a path integral in AF space:
\begin{equation}
\ev*{\hat{H}} = \frac{\int \frac{\mel{\phi_L}{\hat{H}}{\phi_R}}{\ip{\phi_L}{\phi_R}} P(\bm{X}) \braket{\phi_L}{\phi_R} \dd \bm{X}}
{\int P(\bm{X}) \ip{\phi_L}{\phi_R} \dd \bm{X}},
\label{eq:H-Metrop}
\end{equation}
where $\bra{\phi_L} := \bra{\psi_I} \prod_{m = 1}^{M_L} \hat{B}(\bm{x}^{(M - m + 1)})$ and $\ket{\phi_R} := \prod_{m = 1}^{M_R} \hat{B}(\bm{x}^{(m)}) \ket{\psi_I}$. These are single Slater determinants if the initial state is chosen as a single Slater determinant. In the case of a multi-determinant $\ket{\psi_I}$, the different terms in the linear combination can be sampled.
The number $M = \tau / \Delta \tau$ is the total number of time slices, and $M_L$ and $M_R$ correspond to $\tau_L$ and $\tau_R$ respectively.
The variables $\bm{X} := \{\bm{x}^{(1)}, \bm{x}^{(2)}, \ldots, \bm{x}^{(M)}\}$ form a $N_\text{s} \times M$-dimensional vector in AF space.
The probability distribution $P(\bm{X}) := \prod_{m = 1}^M p(\bm{x}^{(m)})$.

We adopt a force bias method~\cite{Hao_PRA2015} instead of the usual single-site heat bath update, which helps reduce the autocorrelation time in the Monte Carlo sampling.
In this method, we update a cluster of AFs at each time slice $n$ simultaneously. The size of the cluster $N_\text{c}$ can be as large as $N_\text{s}$ and can be tuned according to the acceptance ratio.
Below we sketch the algorithm for the spin decomposition; generalization to other HS transformations is straightforward~\cite{Hao_PRA2015,Shiwei_LectureNotes2019}.
Each AF in the cluster is proposed a new value according to
\begin{equation}
\mathcal{P}(x_i) := \frac{\mathrm{e}^{\gamma x_i (\bar{n}_{i \spinup} - \bar{n}_{i \spindown})}}
{\sum_{x_i = \pm 1} \mathrm{e}^{\gamma x_i (\bar{n}_{i \spinup} - \bar{n}_{i \spindown})}},
\label{eq:sample-from-force}
\end{equation}
thus giving a probability density for a new candidate cluster
$\mathcal{P}(\bm{x}^{(n)}) := \prod_{i = 1}^{N_\text{c}} \mathcal{P}(x^{(n)}_i)$.
The optimal choice of force bias is given by
$\bar{n}_{i \sigma} := \mel{\phi_L}{\hat{n}_{i \sigma}}{\phi_R} / \ip{\phi_L}{\phi_R}$, where the $\sigma$ and $i$ are the spin and lattice site indices, and leads to a $\mathcal{P}$ that approximates the target probability to $\mathcal{O}(\sqrt{\Delta \tau})$.

\subsubsection{Doped repulsive Hubbard model: constrained path AFQMC}

All the AFQMC results on the repulsive Hubbard model away from half-filling are obtained with CP-AFQMC~\cite{Shiwei_PRL1995,Shiwei_PRB1997,Shiwei_LectureNotes2019}.
CP-AFQMC is built on an open-ended random walk, with an indefinite value of $\tau$. The sign problem is removed by introducing a trial state $\ket{\psi_T}$ to guide and constrain the random walk. In this work, $\ket{\psi_T}$ is obtained from noninteracting calculations for closed-shell systems and Hartree--Fock calculations for open-shell systems.
In the latter cases, the CP error is further reduced by applying a self-consistent constraint~\cite{Mingpu_selfConsistent_PRB2016} which couples to a generalized Hartree Fock (GHF) calculation with an effective $U_\text{eff}$ determined via the self-consistency~\cite{Mingpu_PRB2016}.
Several results are also provided from constraint release~\cite{Hao_PRB2013} which are essentially exact, as indicated in the main text.

Different from the Metropolis approach discussed above, CP-AFQMC is performed via an open-ended branching random walk, in which a population of walkers $\{\ket*{\phi^{(n)}_k}, w^{(n)_k}\}$ are propagated following the time evolution $\ket*{\psi^{(n + 1)}} = \mathrm{e}^{-\tau \hat{H}} \ket*{\psi^{(n)}}$.
These walkers sample the many-body wave function in the sense that $\ket*{\psi^{(n)}} \propto \sum_k w^{(n)} \ket*{\phi^{(n)}_k} / \ip*{\psi_T}{\phi^{(n)}_k}$.
The ground-state energy is then given by $\ev*{\hat{H}} = \mel{\psi_T}{\hat{H}}{\psi_G} / \ip{\psi_T}{\psi_G}$.
After the random walk has equilibrated (i.e., after a sufficient number of steps $n_\text{eq} = \tau_\text{eq} / \Delta \tau$), the walkers will sample $\ket{\psi_G}$, and all subsequent steps $n > n_\text{eq}$ can be used to compute the ground-state energy:
\begin{equation}
\ev*{\hat{H}} = \frac{\sum_{k, n} \frac{\mel{\psi_T}{\hat{H}}{\phi^{(n)}_k}}{\ip{\psi_T}{\phi^{(n)}_k}}\,w^{(n)}_k}
{\sum_{k, n} w^{(n)}_k}.
\label{eq:H-CP}
\end{equation}

In the random walks, importance sampling is introduced which amounts to sampling new AF to advance the walker by a modified probability density:
\begin{equation}
\ket*{\phi^{(n + 1)}} \gets \int \tilde{p}(\bm{x}) \hat{B}(\bm{x}) \dd \bm{x} \ket*{\phi^{(n)}},
\end{equation}
where $\tilde{p}(\bm{x}) \propto \ip*{\psi_T}{\phi^{(n + 1)}} / \ip*{\psi_T}{\phi^{(n)}}$ builds in the knowledge from $\ket{\psi_T}$ to
improve the sampling efficiency~\cite{Shiwei_LectureNotes2019}. The actual form of $\tilde{p}(\bm{x})$ contains force bias terms similar to how we formulated the generalized Metropolis algorithm above.

The importance sampling transformation also automatically imposes a constraint~\cite{Shiwei_LectureNotes2019}.
As $\Delta \tau$ approaches zero, the random walkers will not cross the surface defined by $\ip*{\psi_T}{\phi^{(n)}_k} = 0$, which separates two degenerate regions in AF space, each of which is overcomplete and can fully represent $\ket{\psi_G}$. Constraining the random walks in one region of the Slater determinant space (or equivalently, the AF space) is an exact condition if $\ket{\psi_T} = \ket{\psi_G}$. The CP approximation uses an approximate $\ket{\psi_T}$ to impose this condition, which leads to a systematic error.
The ground-state energy computed by the mixed estimate of \Eq{H-CP} is therefore \emph{not variational}~\cite{Carlson_PRB1999}.
The CP-AFQMC method has been extensively benchmarked, and the CP error was shown to be small for Hubbard-like systems (see, e.g., Refs.~\cite{Mingpu_PRB2016,LeBlanc_PRX2015,Hao_PRB2013}), and the results provided here for the doped Hubbard model are expected to be very accurate, with the relative error below a few tenths of a percent.

As mentioned, the energies for the doped Hubbard model were computed using either
(i) constraint release (taken from Ref.~\cite{Hao_PRB2013}, essentially exact),
(ii) with free-electron $\ket{\psi_T}$ (for closed-shell systems), or
(iii) with a self-consistent constraint~\cite{Mingpu_selfConsistent_PRB2016}.
In the data included in this paper, we have indicated how each energy was obtained and, in the case of (iii), including the final $U_\text{eff}$ in the GHF after convergence of the self-consistent iteration between GHF and AFQMC.

\subsubsection{Continuous-time quantum Monte Carlo}

The continuous-time quantum Monte Carlo~\cite{mattis2015_241118} algorithm carries out a diagrammatic expansion of the imaginary-time projection operator $\mathrm{e}^{-\tau \hat{H}}$ and samples interaction expansion terms
\begin{align}
& \ev{\mathrm{e}^{-\tau \hat{H}}}{\psi_I} = \nonumber \\
&\phantom{{}\cdot{}} \sum_{k = 0}^\infty (-1)^k \int_0^\tau \dd \tau_1 \left( \prod_{i = 1}^{k - 1} \int_{\tau_i}^\tau \dd \tau_{i + 1} \right) \nonumber \\
&\cdot \ev{\mathrm{e}^{-(\tau - \tau_k) \hat{H}_0} \hat{H}_1 \left( \prod_{i = k - 1}^1 \mathrm{e}^{-(\tau_{i + 1} - \tau_i) \hat{H}_0} \hat{H}_1 \right) \mathrm{e}^{-\tau_1 \hat{H}_0}}{\psi_I},
\label{eq:ct-expansion}
\end{align}
where $\hat{H} = \hat{H}_0 + \hat{H}_1$. Ground state physical observables are measured as
\begin{equation}
\ev*{\hat{O}} = \lim_{\tau \to \infty} \frac{\ev{\mathrm{e}^{-\tau \hat{H} / 2}\,\hat{O}\,\mathrm{e}^{-\tau \hat{H} / 2}}{\psi_I}}
{\ev{\mathrm{e}^{-\tau \hat{H}}}{\psi_I}}.
\end{equation}

For the $t$-$V$ model, the initial state $\ket{\psi_I}$ is chosen as a single Slater determinant, and $\hat{H}_0$ and $\hat{H}_1$ are the non-interaction and the interaction terms respectively. Here, we write $\hat{H}_1 = \frac{V}{4} \sum_{\langle i, j \rangle} \mathrm{e}^{\mathrm{i} \pi (\hat{n}_i + \hat{n}_j)}$ to ensure that each term in the interaction expansion is evaluated as a determinant. Simulation of the spinless $t$-$V$ model on the bipartite lattice at half-filling with repulsive interaction is free from the fermion sign problem~\cite{wanglei2015_235151}. Therefore, the results are free from systematic Trotter or constrained path errors. The computational time complexity of this method scales as $O(\tau\,V N_\text{s}^3)$.

\section{Comparison of the variational methods}

\begin{figure}[htb]
\includegraphics[width=\linewidth]{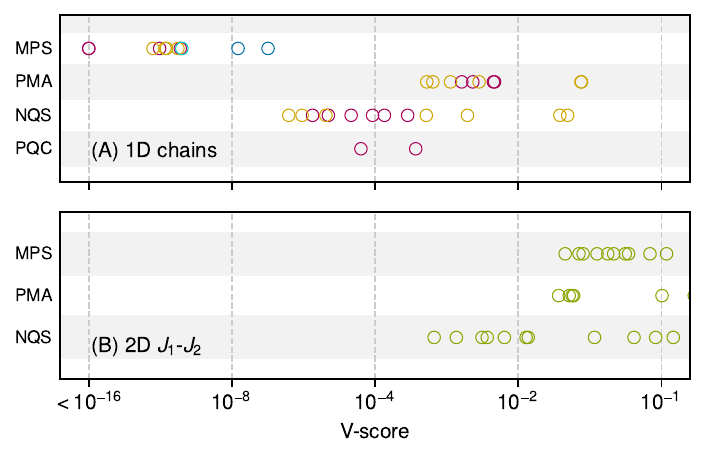}
\caption{\textbf{Comparison of V-scores from different variational methods on given Hamiltonians.}
The Hamiltonian is (\textbf{A}) 1D chain of TFIM (red), Heisenberg (yellow), $t$-$V$ (light blue), or Hubbard (dark blue); and (\textbf{B}) 2D {\JoneJtwo} beyond ED size.
The methods are matrix product states (MPS), VMC with physically motivated ansatzes (PMA), VMC with neural quantum states (NQS), and parameterized quantum circuits (PQC).}
\label{fig:v-score-method}
\end{figure}

In addition to characterizing the hardness of Hamiltonians, the V-score can also serve as a metric to compare the accuracy of different variational methods. As shown in Fig.~{\ref{fig:v-score-method}}, MPS works excellently on 1D problems, while NQS performs better than traditional methods on 2D $J_1$-$J_2$ models that are too large for ED. We expect more research in the future to analyze the performances of different variational methods on a wider range of physical models.

\clearpage


%

\end{document}